# Nonperturbative treatment of the N-body system with dynamical mesons

**C. Alexandrou** and **F. K. Diakonos**

Department of Natural Sciences, University of Cyprus, CY-1678 Nicosia, Cyprus

## Abstract

The variational approach, used by Feynman in the study of the polaron problem, is generalized to treat a system of N non-relativistic particles interacting with scalar and vector mesons. After integrating out the meson fields in the path integral formulation we perform a variational calculation for the effective N-body two-time action to obtain the energy and the mass of the system. The interplay of self energies and exchange terms in attaining binding and saturation is examined. We estimate the size of the particles in the medium and give the mean number of mesons as a function of N.

# 1 Introduction

The description of a strongly interacting many body system starting from the underlying microscopic theory is a fundamental problem in theoretical physics. It is well known that we need a relativistic field theory to describe the fundamental interactions of such an N-body system. However solving for the N-body bound state properties in an exact field theoretic approach is still an unsolved problem. In traditional nuclear physics the approach has been to use in the nonrelativistic many body Schrödinger equation static or near static nucleon-nucleon potentials, deduced from nucleon-nucleon scattering. The solution of the resulting many body Schrödinger equation in most cases involves a number of approximations such as the summation of a selected class of diagrams as in the Brueckner approach. Application of Monte-Carlo techniques, which proved to be extremely useful in the study of single particle properties such as masses of mesons and baryons [1], remains a major problem in the case of many fermions, due to antisymmetry. It is only recently that variational Green's function Monte-Carlo techniques have began to yield a more complete solution to the non-relativistic Schrödinger equation in the case of selected nuclei [2]. Despite the success of the traditional approach we know that it does not present the full picture. We know that the interactions arise from meson exchange leading to non-static potentials, vertex corrections and self energies and that mesons can have a real existence in the nucleus. Therefore one would like to study the many-body system *with* dynamical mesons present. This is the main motivation of the present investigation.

In a previous work [3] the dressing of a single nucleon by mesons was examined using as underlying theory the Walecka model [4]. In this model the main features of the nucleon-nucleon interaction arise from the exchange of scalar and vector mesons. The method used was based on Feynman's variational approach [5] originally applied to the Fröhlich Hamiltonian [6] describing an optical polaron. This approach has been shown by a numerical calculation [7] to be the best semi-analytical method valid for both weak and strong electron-phonon coupling. The success of the method is based on the fact that the retardation effects of the interaction in the original system are taken into account by simulating the system with a retarded harmonic oscillator with properly chosen coupling and retardation constants. In order to apply this approach to the Walecka model we treat the nucleons, being heavier than the mesons, nonrelativistically and keep only terms of order $1/M$ where $M$ is the nucleon mass. To this order there is also a spin-dependent part which we neglect for technical reasons. Within the nonrelativistic approximation the Walecka model resembles closely the Fröhlich Hamiltonian. In the path integral formulation the mesonic degrees of freedom can be integrated out in an exact manner giving rise to a non local (two-time) effective N-body action.

The properties of the N-body system are evaluated using Feynman's variational ansatz generalized for the N-body action. In this way we sum up self energy and exchanged diagrams to all orders, albeit in an approximate way, with only the first order perturbative results reproduced exactly for the energy and effective mass. In addition to the binding energy and mass of the system we calculate the size of the dressed constituent in the medium and the excess number of scalar and vector mesons in the bound N-body state, both of which are interesting quantities in connection with the various explanations of the EMC effect [8]. The results obtained throughout this work should only be taken as a qualitative description of the corresponding quantities of real nuclei. The reason for this is that, in addition to the nonrelativistic approximation and to neglecting the spin dependence, we have not imposed antisymmetry. As we will explain in section 3 imposing antisymmetry destroys the quadratic form of the trial action and one can no longer perform the functional integrations analytically, a prerequisite for the present approach. Thus the strong repulsive short range interaction is the only factor that prevents two particles overlapping with each other. We expect the effective "hard" core to be softer in comparison to the real nuclear case. Nevertheless the bound states approximate a saturating system for special values of the coupling constants and we will still refer to the constituents as nucleons.

The paper is organized as follows: In section 2 we derive the many body action for the Walecka model after elimination of the mesonic degrees of freedom and discuss the static two-body potential. In section 3 we calculate the energy and the effective mass of the multiparticle state. We derive also expressions for the mean number of scalar and vector mesons and for the radius of the bound state. In section 4 we present our numerical results and discuss the stability of the N-particle system. The details of the evaluation of the various observables can be found in the appendices.

## 2  Effective many-body action

The interaction part of the Langrangian in the Walecka model consists of two contributions: i) an attractive interaction between the nucleons ($\psi$) due to the exchange of scalar meson fields ($\phi$) which is responsible for the binding of the nucleons and, ii) a repulsive short range term which is described through the exchange of vector meson fields ($V_\mu$) and is responsible for the stability of the multi-nucleon bound state. In terms of the nucleon and meson fields the Langrangian can be written as

$$\mathcal{L} = \mathcal{L}_0(\psi, \phi, V_\mu) + g_S \bar{\psi}\psi\phi - g_V \bar{\psi}\gamma_\mu\psi V^\mu \tag{2.1}$$

where the free part is given by

$$\mathcal{L}_0(\psi, \phi, V_\mu) = \bar{\psi}(i\gamma_\mu\partial^\mu - M)\psi + \frac{1}{2}(\partial_\mu\phi\partial^\mu\phi - m_S^2\phi^2) - \frac{1}{4}F_{\mu\nu}F^{\mu\nu} + \frac{1}{2}m_V^2 V_\mu V^\mu \quad . \tag{2.2}$$

To obtain the effective action for the interacting nucleons we have to perform the path integral over the meson fields. There are no interaction terms between the scalar and the vector mesons in the above Langrangian and therefore the integration can be easily done leading to an effective action with two interaction terms which correspond to the scalar and vector contribution respectively. The exact form obtained is given by

$$S = S_0 + \frac{1}{2}\int d^4x d^4x'[-g_S^2 \rho_S(x) G_S(x-x') \rho_S(x') + g_V^2 J_\mu(x) G_V^{\mu\nu}(x-x') J_\nu(x')] \tag{2.3}$$

where $S_0$ is the kinetic term, $\rho_S(x) = \bar{\psi}(x)\psi(x)$ is the scalar density and $J_\mu = \bar{\psi}(x)\gamma_\mu\psi(x)$ is the baryon current. $G_V^{\mu\nu}$ and $G_S$ are the propagators of the vector and scalar mesons respectively. The scalar propagator $G_S$ is given by

$$G_S(x-x') = \int \frac{d^4k}{(2\pi)^4} \frac{i\, e^{-ik.(x-x')}}{k^2 - m_S^2 + i\epsilon} \tag{2.4}$$

and the corresponding expression for the massive vector field propagator, $G_V^{\mu\nu}$, reads

$$G_V^{\mu\nu}(x-x') = \int \frac{d^4k}{(2\pi)^4} \frac{i(g^{\mu\nu} - k^\mu k^\nu/m_V^2)}{k^2 - m_V^2 + i\epsilon} e^{-ik.(x-x')} \quad . \tag{2.5}$$

It is well known [9] that, due to the $k^\mu k^\nu$ term, the vector meson propagator develops a singular ultraviolet behaviour. However the baryonic current is conserved and we can therefore neglect this term, since, in every physical quantity, it always occurs contracted with a current of vanishing divergence [4].

In order to stay as close as possible to the Feynman ansatz the main assumption that we make is to consider the nucleons much heavier than the mesons so that we can treat them *non-relativistically*. We realize the non-relativistic approximation by going over to the particle picture for the nucleons keeping terms of $\mathcal{O}(1/M)$. The scalar density is then given by

$$\rho_S(x) \simeq \psi^\dagger(x)\psi(x) = \sum_{i=1}^N \delta(\mathbf{x} - \mathbf{x}_i) \tag{2.6}$$

and the vector current [10] by

$$J^\mu(x) = \sum_{i=1}^N J_i^\mu(x) \ , \qquad J_i^\mu(x) = \frac{dx_i^\mu}{dt} \delta^3(\mathbf{x} - \mathbf{x}_i(t)) + \text{spin current} \ , \ x_1^0 = x_2^0 = ... = x_N^0 = t \tag{2.7}$$

The above current is conserved, i.e. it fulfills the continuity equation. The non-relativistic approximation is incorporated in the use of a common time variable for all nucleon trajectories. The spin dependent part in eq. (2.7) will be neglected in what follows in order to avoid the complications of spin in a path integral [11]. Inserting the non-relativistic current into eq. (2.3) and performing the $k_0$ integration we get for the effective action

$$\begin{aligned} S_{\text{eff}} &= \sum_{i=1}^N \int_0^\beta dt \left\{ \frac{M}{2}\dot{\mathbf{x}}_i^2 - \sum_{j=1}^N \int \frac{d^3k}{8\pi^2} \int_0^\beta dt' e^{i\mathbf{k}\cdot(\mathbf{x}_i(t)-\mathbf{x}_j(t'))} \left[ R_S(k,t-t') + R_V(k,t-t') \right. \right. \\ &\quad \cdot \left. \left( \frac{m_V^2}{k^2} - \dot{\mathbf{x}}_{\perp(i)}(t) \cdot \dot{\mathbf{x}}_{\perp(j)}(t') \right) \right] + \sum_{j=1}^N \int \frac{d^3k}{4\pi^2} \omega_V(k) \frac{1}{k^2} R_V(k,0) e^{i\mathbf{k}\cdot(\mathbf{x}_i(t)-\mathbf{x}_j(t))} \right\} \end{aligned} \tag{2.8}$$

with

$$R_i(k,\tau) = \frac{g_i^2}{4\pi} \frac{F_i^2(k)}{\omega_i(k)} \exp(-\omega_i(k)|\tau|) \quad \omega_i(k) = \sqrt{k^2 + m_i^2} \quad i = S, V \ . \tag{2.9}$$

We stress that up to this point the mesonic degrees of freedom are treated in an exact fully relativistic manner. Considering the Walecka model as a low energy effective theory we have introduced form factors $F_i(k)$ at the meson-nucleon vertices to cut off the high-momentum components. Here we use a standard monopole-type form factor [12]

$$F_i(k) = \frac{\Lambda_i^2}{\Lambda_i^2 + k^2}, \qquad i = S, V \tag{2.10}$$

In deriving eq.(2.8) we have *not* respected the Pauli principle for the nucleons for purely technical reasons. To project onto the antisymmetric space within the present formulation one uses antisymmetric coordinate states [13] to obtain a determinant in the particle labels. The simplest way to see this is to consider a free gas of fermions of mass $m$. Then the infinitesimal evolution operator at the $k^{\text{th}}$ time slice between antisymmetric coordinate states is given by

$$< \mathbf{x}_1(k), ..., \mathbf{x}_N(k)|e^{-\epsilon H}|\mathbf{x}_1(k-1), ..., \mathbf{x}_N(k-1) > = \left(\frac{m}{2\pi\epsilon}\right)^{N/2} DetM(k) \; e^{-\frac{m}{2\epsilon}\sum_i (\mathbf{x}_i(k)-\mathbf{x}_i(k-1))^2} \tag{2.11}$$

where

$$M_{ij}(k) = e^{-\frac{m}{2\epsilon}\left[(\mathbf{x}_j(k)-\mathbf{x}_i(k-1))^2 - (\mathbf{x}_i(k)-\mathbf{x}_i(k-1))^2\right]}$$

and $\epsilon$ is the infinitesimal time step. Therefore including antisymmetry on the level of the trial action introduces determinantal dependences on the particle trajectories and we are no longer able to perform the functional integrals analytically destroying one of the criteria required by our method. There are attempts to model the Pauli principle classically [14] and one could envisage applying those ideas within the present approach. Progress in this direction will be reported elsewhere.

This effective action has several important features:

i) Due to the interaction with relativistic mesons the $\mathbf{k}$-dependence of the action is more complicated as compared to the polaron action where integration over the $1/k$ form factor gives rise to the retardated Coulomb potential (more on this point in appendix A).

ii) The scalar mesons give rise to a two time effective action which closely resembles the polaron action for $N = 1$.

iii) The vector mesons generate in addition a transverse velocity dependent contribution to the two time effective action which is absent in the Fröhlich Hamiltonian. For $N > 1$ they also produce an equal-time exchanged term.

The static two-body potential can be easily obtained from $S_{\text{eff}}$ in the following manner: We make the transformation

$$T = \frac{t+t'}{2} \qquad \tau = t - t' \tag{2.12}$$

and we consider the limit $\tau \to 0$ so that $\mathbf{x}_i(t) - \mathbf{x}_j(t') = \mathbf{x}_i(T+\frac{\tau}{2}) - \mathbf{x}_j(T-\frac{\tau}{2}) \sim \mathbf{x}_i(T) - \mathbf{x}_j(T)$. In the absence of the form factors and using the integration formulas of the modified Bessel functions given in appendix A we can perform the $\tau$ integration. We find the familiar sum of a repulsive and an attractive Yukawa potential, typical for vector and scalar massive meson exchange

$$V(\mathbf{r}) = \frac{g_V^2}{4\pi}\frac{e^{-m_V|\mathbf{r}|}}{|\mathbf{r}|} - \frac{g_S^2}{4\pi}\frac{e^{-m_S|\mathbf{r}|}}{|\mathbf{r}|} \tag{2.13}$$

where $\mathbf{r} = \mathbf{x}_i - \mathbf{x}_j$. To appreciate the modifications introduced by the presence of the form factors we give the static potential for the monopole-type form factors used. It reads

$$V(\mathbf{r}) = \alpha_V \left\{ \frac{\Lambda_V^4}{(\Lambda_V^2 - m_V^2)^2} \frac{(e^{-m_V|\mathbf{r}|} - e^{-\Lambda_V|\mathbf{r}|})}{|\mathbf{r}|} - \frac{\Lambda_V^3}{2(\Lambda_V^2 - m_V^2)} e^{-\Lambda_V|\mathbf{r}|} \right\}$$

$$- \quad \alpha_S \left\{ \frac{\Lambda_S^4}{(\Lambda_S^2 - m_S^2)^2} \frac{(e^{-m_S|\mathbf{r}|} - e^{-\Lambda_S|\mathbf{r}|})}{|\mathbf{r}|} - \frac{\Lambda_S^3}{2(\Lambda_S^2 - m_S^2)} e^{-\Lambda_S|\mathbf{r}|} \right\} \tag{2.14}$$

where

$$\alpha_i = \frac{g_i^2}{4\pi} \quad i = S, V \ . \tag{2.15}$$

If the cut off $\Lambda_S$ and $\Lambda_V$ are large compared to the meson masses $m_S$ and $m_V$ then we obtain the result of (2.13). From fits to nuclear properties the values obtained for the meson masses are

$$m_S = 520 \, \text{MeV} \qquad\qquad m_V = 780 \, \text{MeV} \ . \tag{2.16}$$

Since we have treated the Walecka model as a low energy effective theory for the nucleons the cut offs should be of the order of the nucleon mass in order to eliminate large momenta. This requirement in combination with the fact that the cut offs should be larger than the meson masses makes the non-relativistic approximation not very well fulfilled in this model. It would be more appropriate to study the pion-nucleon system within this approach if it were not for the complications of spin and isospin. We show the form of this potential for various values of the cutoff parameters $\Lambda_S, \Lambda_V$ in fig.1. As it can be seen, for this allowed range, the effects of the cut off are rather strong altering the height of the repulsive core and the depth of the minimum.

## 3  Trial action and observables

A good trial action must incorporate the essential physics contained in the effective theory as given by (2.8). At the same time it must be simple enough so that we can perform analytically the functional integrals necessary to calculate our observables. This gives us the following conditions for the trial action:

i) It must be translationally invariant.

ii) It must include the effects of retardation.

iii) It can only be at the most a quadratic function of the particle trajectories.

The last criterion will not be fulfilled if antisymmetry is included. To construct a suitable trial action we follow Feynman's approach and represent the mesons by fictitious particles harmonically coupled to the nucleons. To take into account the self energies we attach a fictitious particle by a spring to each nucleon. We will refer to this particle as the $y-$particle. The exchange of mesons is modeled by attaching all nucleons to an additional fictitious particle, the $Z-$particle. This system is schematically shown in fig. 2 and corresponds to the Langragian

$$L = \sum_{i=1}^{N} \left[ \frac{M}{2} \dot{\mathbf{x}}_i^2 + \frac{M_Z}{2} \dot{\mathbf{Z}}^2 + \frac{m_y}{2} \dot{\mathbf{y}}_i^2 - \frac{K}{2} (\mathbf{Z} - \mathbf{x}_i)^2 - \frac{\kappa}{2} (\mathbf{y}_i - \mathbf{x}_i)^2 \right] \tag{3.17}$$

Integrating out the fictitious particles in the Feynman path integral we obtain after rotation to Euclidean time and taking the $\beta \to \infty$ limit

$$S_F = \int_0^\beta dt \sum_{i=1}^N \left\{ \frac{M}{2}\dot{\mathbf{x}}_i^2 + C_D \int_0^\beta dt' (\mathbf{x}_i(t) - \mathbf{x}_i(t'))^2 e^{-w_D|t-t'|} \right.$$
$$\left. + C \sum_{j=1}^N \int_0^\beta dt' (\mathbf{x}_i(t) - \mathbf{x}_j(t'))^2 e^{-w|t-t'|} \right\} , \qquad (3.18)$$

with

$$\omega_D = \sqrt{\frac{\kappa}{m_y}}, \qquad C_D = \frac{\omega^3 m_y}{4} \quad ; \quad \omega = \sqrt{\frac{NK}{M_Z}}, \qquad C = \frac{\omega^3 M_Z}{4N^2} . \qquad (3.19)$$

Since the above action is quadratic in $\mathbf{x}_i$ all the path integrals can be done analytically. However the resulting expressions for the various observables are rather involved and are given in appendix B. Instead we consider here a special case of the above trial action derived from the Lagrangian of (3.17) by setting $\kappa = 0$; in other words we model both exchange and self-energy contributions by a single fictitious particle. It corresponds to setting $C_D = 0$ in (3.18) so that exchange and direct terms have the same strength and retardation reducing the number of variational parameters to two. It is the simplest extension of Feynman's trial action for the many body system and we will refer to it in what follows as trial action 1. This trial action is used mainly as an illustration of the main steps of our calculation. In order to demonstrate that it is important to have different strengths for the diagonal and non-diagonal terms in the trial action irrespective of the retardation we consider the special case of allowing different $C$ and $C_D$ but still keeping the same retardation. It corresponds to a model Lagrangian of the system shown in fig. 2 but with $\kappa/m_y = NK/M_Z$ and it emerges from (3.18) if we set $w = w_D$. We will refer to this 3-parameter action as trial action 2. It allows us to investigate the relative strength of diagonal and exchange terms of the trial action. This separation is lost in the four parameter trial action (trial action 3) given by (3.18) since, in addition to the purely diagonal terms with retardation $w_D$, there are diagonal terms having retardation factor and strength the same as that of the exchange. A way to recover this separation is to exclude diagonal terms in the sum of the exchange terms in $S_F$, leading to trial action 4. This will be useful when we discuss the dressing of the constituent which is the main reason of introducing trial action 4. In contrast to the rest of the trial actions it can no longer be derived from an underlying model Lagrangian. Trial action 4 completes the variational ansätze that we considered in this work.

A variational bound on the energy is obtained on the level of an action by using Jensen's inequality,

$$\frac{\int \mathcal{D}x(t) e^{-S} e^{-S_F}}{\int \mathcal{D}x(t) e^{-S_F}} \equiv <e^{-S}> \geq e^{-<S>} . \qquad (3.20)$$

This gives for the energy $E_N$ of the multi-particle system the upper limit

$$E_N \leq E_F + \lim_{\beta \to \infty} \frac{<S_{\text{eff}} - S_F>}{\beta} \qquad (3.21)$$

where $E_F$ is the energy corresponding to the trial action $S_F$ and the expectation value is taken with respect to the trial action $S_F$.

In order to obtain the effective mass $M_{\text{eff}}(N)$ of the $N$-particle system we calculate the contribution to the energy by giving the center of mass mean velocity $\mathbf{u}_{\text{cm}}$. The trajectories of the nucleons now satisfy the boundary condition $\mathbf{x_i}(\beta) = \mathbf{x_i}(0) + \mathbf{u}_{\text{cm}}\beta$. Using the calculated energy $E_N(\mathbf{u}_{\text{cm}})$ we obtain the effective mass of the system from

$$M_{\text{eff}}(N) = \nabla^2 \, E_N(\mathbf{u}_{\text{cm}})|_{u_{\text{cm}}=0} \tag{3.22}$$

The path integrals involved in the computation of the expectation value in eq. (3.21) can be done easier using a Fourier representation for the particle paths with the path of the ith particle given by

$$\mathbf{x}_i(t) = \frac{1}{2}\mathbf{a}_i^0 + \sum_{n=1}^{\infty} \mathbf{a}_i^n \cos(2\pi n \frac{t}{\beta}) + \sum_{n=1}^{\infty} \mathbf{b}_i^n \sin(2\pi n \frac{t}{\beta}) \quad . \tag{3.23}$$

As long as the particle trajectories occur at most quadratically in the exponent of the path integral the stationary phase method is exact and easy to apply. The details of the calculation can be found in appendix B. There we also give the results for the energy, the effective mass and the other observables using trial action 3 ( and consequently 2) and 4. Because of the complexity of these expressions we will discuss here the results obtained using trial action 1. For the energy we find

$$\begin{aligned}E_N = &\; E_F - \frac{1}{\pi} \int_0^\beta d\tau \int_0^\infty dk\, k^2 \left\{ N R_S(k,\tau) \left( (N-1)e^{-\frac{k^2}{2M}\mu^2(\tau)} + e^{-\frac{k^2}{2M}\mu_D^2(\tau)} \right) + N R_V(k,\tau) \right. \\ & \left. \cdot \left( (N-1) \left[ \frac{m_V^2}{k^2} + \frac{c^2 e^{-v\tau} - c\, v\, e^{-c\tau}}{MNv} \right] e^{-\frac{k^2}{2M}\mu^2(\tau)} + \left[ \frac{m_V^2}{k^2} + \frac{c^2 e^{-v\tau} + (N-1)c\, v\, e^{-c\tau}}{MNv} \right] e^{-\frac{k^2}{2M}\mu_D^2(\tau)} \right) \right\} \\ & + \frac{1}{\pi} \int_0^\infty dk\; N(N-1)\; R_V(k,0)\; \omega_V\; e^{-\frac{k^2}{2Mc}} + C_0 \end{aligned} \tag{3.24}$$

where we have used the notation

$$c = \sqrt{\frac{4CN}{Mw}}, \qquad v^2 = c^2 + w^2 \quad . \tag{3.25}$$

$E_F$ is the energy corresponding to the trial action 1 and is given by

$$E_F = \frac{3}{4}\frac{(v-w)^2}{v} + (N-1)\,c \quad . \tag{3.26}$$

The quantities $\mu^2(\tau)$ and $\mu_D^2(\tau)$ are given by

$$\begin{aligned}\mu^2(\tau) &= \frac{1}{N}\left(\frac{v^2-w^2}{v^3}(1-e^{-v\tau}) + \frac{w^2}{v^2}\tau + \frac{1}{c}(N-1+e^{-c\tau})\right) \\ \mu_D^2(\tau) &= \frac{1}{N}\left(\frac{v^2-w^2}{v^3}(1-e^{-v\tau}) + \frac{w^2}{v^2}\tau + \frac{(N-1)}{c}(1-e^{-c\tau})\right)\end{aligned} \tag{3.27}$$

and $C_0$, which is independent of the variational parameters, by

$$C_0 = \frac{N}{\pi} \int_0^\infty dk\; k^2\; R_V(k,0)\left(\frac{1}{M} + \frac{\omega_V}{k^2}\right) \quad . \tag{3.28}$$

For the effective mass we find

$$M_{\text{eff}}(N) = MN \left( 1 \; + \; \frac{2}{3\pi} \int_0^\infty d\tau \int_0^\infty dk \, k^2 \left\{ \frac{k^2 \tau^2}{2M} \, R_S(k,\tau) \left[ (N-1) e^{-\frac{k^2}{2M}\mu^2(\tau)} + e^{-\frac{k^2}{2M}\mu_D^2(\tau)} \right] \right. \right.$$
$$+ \; \frac{k^2 \tau^2}{2M} \, R_V(k,\tau) \left[ (N-1) \left( \frac{m_V^2}{k^2} + \frac{c^2 e^{-v\tau} - c \, v \, e^{-c\tau}}{MNv} \right) e^{-\frac{k^2}{2M}\mu^2(\tau)} \right.$$
$$+ \; \left. \left( \frac{m_V^2}{k^2} + \frac{c^2 e^{-v\tau} + (N-1) c \, v \, e^{-c\tau}}{MNv} \right) e^{-\frac{k^2}{2M}\mu_D^2(\tau)} \right]$$
$$+ \; \left. \left. \frac{2}{M} R_V(k,0) \left[ (N-1) e^{-\frac{k^2}{2M}\mu^2(\tau)} + e^{-\frac{k^2}{2M}\mu_D^2(\tau)} \right] \right\} \right) \; . \tag{3.29}$$

A check of our variational calculation is the reduction of our expressions to those obtained in perturbation theory when $g_V \to 0$ and $g_S \to 0$ which in our formalism means $v \to w$. Taking $v = w + \mathcal{O}(g^2)$ in the above expressions the contribution from the non-diagonal part of the action vanishes and we obtain the sum of the single particle self-energies i.e. no binding which is the perturbative result.

Other interesting parameters of the many particle system are the average interparticle distance and the mean number of mesons present in a bound state. The mean number of scalar and vector mesons is obtained by finding the expectation value of the corresponding number operator to zeroth order i.e. taking the expectation value with respect to the trial action. The method is the same as for the one particle case described in [3] and therefore we only give here the resulting expressions. We find for the mean number of scalar mesons

$$< N_S > = \frac{N}{\pi} \int_0^\infty dk \int_0^\beta d\tau \, k^2 \tau R_S(k,\tau) \left[ (N-1) e^{-\frac{k^2}{2M}\mu^2(\tau)} + e^{-\frac{k^2}{2M}\mu_D^2(\tau)} \right] \tag{3.30}$$

and of vector mesons

$$< N_V > = \frac{N}{\pi} \int_0^\infty dk \int_0^\beta d\tau \, k^2 \tau R_V(k,\tau) \left\{ (N-1) \, e^{-\frac{k^2}{2M}\mu^2(\tau)} \right.$$
$$\cdot \left[ \frac{k^2}{m_V^2} \left( 1 + \frac{\omega_V}{2M} \frac{d\mu^2(\tau)}{d\tau} \right)^2 - \frac{1}{M} \frac{d^2\mu^2(\tau)}{d\tau^2} \left( 1 + \frac{\omega_V^2}{2m_V^2} \right) \right]$$
$$+ \left. \left[ \frac{k^2}{m_V^2} \left( 1 + \frac{\omega_V}{2M} \frac{d\mu_D^2(\tau)}{d\tau} \right)^2 - \frac{1}{M} \frac{d^2\mu_D^2(\tau)}{d\tau^2} \left( 1 + \frac{\omega_V^2}{2m_V^2} \right) \right] e^{-\frac{k^2}{2M}\mu_D^2(\tau)} \right\} \; . \tag{3.31}$$

The above formulae hold for a general quadratic trial action and for $N = 1$ reduce to those given in ref. [3]. The explicit expressions for $\mu_D^2(\tau)$, $\mu^2(\tau)$ and their derivatives depend of course on the specific trial action.

The interparticle distance is defined by

$$\bar{r}_N^2 = \frac{1}{N(N-1)} \sum_{i,j=1}^N < (\mathbf{x}_i - \mathbf{x}_j)^2 > \tag{3.32}$$

and for a system showing saturation i.e. constant particle density the size of the system is related to the interparticle distance by

$$R_N = \bar{r}_N N^{1/3} \; . \tag{3.33}$$

We evaluate the interparticle distance to zeroth order. The simplest way to do this is to use the wave function corresponding to the variational Hamiltonian. For the simplest trial Hamiltonian we have harmonic couplings of the nucleons to one fictitious particle. In this case the intrinsic wave function is given by

$$\psi(\mathbf{x}_1, ..., \mathbf{x}_N, \mathbf{Z}) = \mathcal{N} \exp\left[-\alpha \sum_{i=1}^{N}(\mathbf{Z} - \mathbf{x}_i)^2 - \frac{\beta}{2} \sum_{i,j=1}^{N}(\mathbf{x}_i - \mathbf{x}_j)^2\right] \quad (3.34)$$

where

$$\alpha = \frac{1}{2}\sqrt{KM_{\text{red}}} \qquad \beta = \frac{1}{N}\left(\sqrt{\frac{KM}{4}} - \alpha\right) , \quad (3.35)$$

$M_{\text{red}} = \frac{MM_Z}{M+M_Z}$ is the reduced mass and $\mathcal{N}$ is the normalisation constant. In addition to the interparticle distance a quantity of interest is the size of the particle *in the medium*. If we were to probe a nucleon with a meson the nucleon - meson vertex will be dressed giving rise to a form factor. The size thus measured can be considered as the core radius of the nucleon. In the one particle case this definition was shown [3] to be identical in zeroth order to the nucleon root mean square radius calculated using the intrinsic wave function of eq. (3.34). We use the same definition in this many particle case where we consider the fictitious and the ith particles as an independent subsystem. Relative to the center of mass of this subsystem we obtain for the root mean square radius or center of mass radius, $r_{\text{cm}}$, of the nucleon in the medium

$$r_{\text{cm}}^2 = \frac{1}{N}\sum_{i=1}^{N} < \left(\frac{M_Z \mathbf{Z} + M\mathbf{x}_i}{M_Z + M} - \mathbf{x}_i\right)^2 > = \left(\frac{N(v^2 - w^2)}{N(v^2 - w^2) + w^2}\right)^2 r_{\text{cloud}}^2 . \quad (3.36)$$

The cloud radius, $r_{\text{cloud}}$, is a measure of the meson cloud around the bare nucleon and in the one-particle case it is defined as the distance of a particle from the fictitious particle. The generalization

$$\bar{r}_{cloud}^2(N) = \frac{1}{N}\sum_{i=1}^{N} < (\mathbf{Z} - \mathbf{x}_i)^2 > = \frac{3}{2NM}\left\{\frac{v}{v^2 - w^2} + \frac{N-1}{\sqrt{v^2 - w^2}}\right\} \quad (3.37)$$

for the N-particle system includes, in addition to the self energies, exchange contributions and it is therefore not the right quantity to measure the size of a single nucleon. Unfortunately for this simple trial action we have no alternative. For our other trial actions the $y$-particle describes purely self energies and therefore the distance of a particle from its $y$-particle is an unambiguous contribution to the cloud radius. We will discuss this point further when we present our results. As in the one particle case the cloud radius for this simple trial action is always larger than the rms (core) radius approaching each other in the strong coupling limit. For $N \gg 1$ the cloud radius approaches the limit

$$r_{\text{cloud}}^2 \xrightarrow{N \gg 1} \frac{3}{2M}\frac{1}{\sqrt{v^2 - w^2}} \quad (3.38)$$

which is equal to the interparticle distance. This is to be expected since the definition given in (3.37) includes the exchanges which probe the interparticle distance. We remark that what we call here cloud radius is in agreement for $N = 1$ with the polaron radius as defined in [3].

All the corresponding formulae for the above quantities in the case of the 3- and 4-parameter trial actions are given in Appendix B.

# 4  Results and Discussion

As it was mentioned in the introduction we are interested in the *qualitative* description of the nuclear many body system mainly due to the lack of antisymmetry. We are thus addressing questions like what is the relative importance of self energies and vertex corrections in a saturating bound system? How do these change as we go to a strongly bound system? What is the mean number of mesons and the amount of dressing in a saturating system and how are they modified as the binding is increased? The purpose is *not* to reproduce the binding energy of real nuclei nor to describe accurately the Deuteron where one would need a much more sophisticated calculation. Therefore our approach in this work is not to search the whole parameter space in order to fix the scalar and vector couplings and form factors to reproduce the binding energies of real nuclei. Instead we consider four sets of these parameters that we regard representative with respect to the posed questions. The values of the couplings $g_S$ and $g_V$ entering the Walecka model, and of the cutoffs $\Lambda_S$ and $\Lambda_V$ are listed in table 1 and we denote these sets by $G_1, G_2, G_3$ and $G_4$. $G_0$ in table 1 is a set of parameters taken from ref. [15] and resulted from a fit to nuclear properties in a mean field approximation. $G_0$ has no form factors and leads to divergent loop integrals. We therefore introduce a scalar and vector form factor each with a cut off. Following ref. [3] we fixed the bare mass of the nucleon $M$ by requiring that the mass of a single "nucleon" be equal to the physical nucleon mass i.e. we have set the effective mass $M_{\text{eff}}(1) = 938.9$ MeV. As was pointed in [3] increasing the value of the form factors decreases the bare mass to an extent that our nonrelativistic approximation can no longer be justified. Therefore the values of the cut offs are restricted for a given set of couplings. For large couplings the cut offs become close to the meson masses. Decreasing the couplings allows an increase in the values of the cut offs. With these considerations we compromised with the sets listed in table 1. There we also give the values for the bare mass and other single particle observables. For $N = 1$ trial action 1, 2 and 4 coincide and give the results of table 1 whereas trial action 3 produces very similar values [1]. The two first sets with the stronger couplings reduce the bare mass by more than 600 MeV and are at the limit of applicability of the non-relativistic nucleon dynamics. $G_3$ and $G_4$ on the other hand produce reasonable bare masses.

The form of the static two-body potential corresponding to these sets of parameters is presented in fig.1. Including the form factors for the couplings of $G_0$ softens the core and reduces the depth of the potential. As we can see in fig. 1 the maximum attraction is the same for the two sets $G_1$ and $G_2$. The repulsive core progressively decreases as we go from $G_1$ to $G_4$. The ratio of maximum attraction to maximum repulsion increases from 0.03 for $G_1$ to 0.15 for $G_3$ to the large value of 1.8 for $G_4$.

Having fixed the bare mass we determine the variational parameters by minimizing the energy $E_N$ of the N-particle state [2]. The dependence on $N$ of the variational parameters obtained in the minimization of the N-particle energies is displayed in fig. 3 for all trial actions and for the parameter sets $G_2$ and $G_4$ using lines for the former and symbols for the latter [3]. The parameters $w_D$ and $v_D$ of the purely diagonal terms in $S_F$ are much larger than $v$ and $w$ and therefore $w_D$ and $v_D$ are scaled by 0.1 in order to display them on the same figure. For action 2 we have set $w = w_D$ and since $v \geq w$ all parameters are scaled by 0.1. For set $G_2$ the dependence of all the variational parameters on $N$ is very weak and although the strength $C$ of the exchange term is small, including this term in the action lowers the energy. For the set $G_4$ the $N$ dependence of $w$ and $v$ is stronger and the strength $C$ becomes larger, in agreement with the fact that the exchange terms dominate the behaviour of a strongly bound system. Having determined the variational parameters we proceed to discuss the various observables.

---

[1] Using trial action 3 decreases the values of the bare mass by 12 MeV in the case of parameter sets $G_1$ and $G_2$ and by 0.4 MeV for $G_3$ and $G_4$.

[2] The minimization was performed with the CERN subroutine MINUIT while the expressions for the energy and the effective mass were evaluated numerically with typically 64 Gauss points.

In table 2 we list the energies obtained from the minimization using the four trial actions for the parameter set $G_2$. The biggest improvement comes from including a coupling of each nucleon to a fictitious particle modeling the self energy which trial action 1 lacks [4]. The variational bound is lower by a significant amount by using trial action 2 indicating the importance of having different strengths $C$ and $C_D$ for the direct and exchanged terms even though the retardation is the same. Allowing in addition a different retardation for the exchange terms as is done for trial action 3 and 4 lowers the energy by a smaller amount. The fact that trial action 3 produces the best bound, also compared to trial action 4, indicates the importance of the extra diagonal term of strength the same as that of the exchange terms a feature shared by the effective action, $S_{\text{eff}}$.

Since the one-particle energies are the same for trial actions 1,2 and 4 the variational principle applies on the level of their binding energies per particle defined as

$$BE_N = -\frac{E_N - NE_1}{N} \quad . \tag{4.39}$$

The results for the binding energy per particle as a function of $N$ are summarized in fig. 4 for the four different sets of parameters and trial actions. For the set $G_1$ the system exhibits an almost saturating behaviour but, for these values of the couplings and with no antisymmetry, with an order of magnitude larger $BE_N$ as compared to the nuclear system. For the rest of the parameter sets the slope becomes increasing larger with $G_4$ as an extreme case where $BE_N$ grows rapidly. The main observation here is that whereas trial action 1 gives no binding for the sets with the larger repulsion ($G_1$, $G_2$) it yields progressively better results as we go to $G_3$ and $G_4$. Similarly results from trial actions 3 and 4 become progressively more similar as we go to $G_3$ and $G_4$, whereas results using trial action 2 always coincide with those from trial action 4. Therefore we conclude that for a saturating system it is very important to construct a good trial action whereas for a strongly bound system this becomes less crucial. The fact that trial action 2 yields results the same as 4 indicates that it is more important to allow a different strength for the diagonal and non-diagonal terms rather than a different retardation. Having compared the various trial actions we proceed to discuss the results obtained using trial action 3 which is superior to the rest.

It is interesting to look separately at the diagonal and non-diagonal contributions to the total binding energy. These are shown in fig. 5 for all four sets of parameters using trial action 3. For $G_1$ and $G_2$ we observe a large cancelation between the diagonal and exchanged contributions. As the repulsive core is decreased the slope of the exchanged contribution increases changing from negative to positive at a decreasing value of $N$. The diagonal contribution remains constant and therefore the $N$ dependence of $BE_N$ is determined by the behaviour of the exchanged part which becomes dominant for the set $G_4$.

---

[3]The behaviour of the variational parameters for the sets $G_1$ and $G_3$ is in overall similar to that of $G_2$.

[4]Allowing for two different exponentials (4-parameters) in trial action 1, which corresponds to coupling all nucleons to two fictitious $Z-$ particles, leads to a minor improvement of the variational bound for $N > 1$.

We can further break up the total diagonal and non-diagonal contributions to those coming from $\Delta S = \frac{1}{\beta} < S_{\text{eff}} - S_F >$ and those coming from the energy, $E_F$, corresponding to the trial action. These separate contributions to the energy and binding energy per nucleon are plotted in fig. 6 for the set $G_1$ and $G_4$ using again trial action 3. We notice that in both cases the diagonal part of $E_F$ is small and constant and therefore the contribution of this term to $BE_N$ is just its single particle value. In the case of $G_1$ the rest of the terms grow linearly with $N$ and therefore give a constant contribution of the same order of magnitude to $BE_N$ [5]. Going from the set $G_1$ to $G_2$ we begin to see a non-linear dependence of the non-diagonal part of $\Delta S$ which leads to the slope of $BE_N$. This observation is independent of the trial action and is due to the increase of the binding in our model. When we go to $G_4$ the non-diagonal $\Delta S$ piece has a significant slope and becomes the dominant contribution. For this last set all trial actions produce results in much closer resemblance to those displayed in fig. 6.

The effective mass defined by (3.22) is a measure of the kinetic mass of the $N$-particle state which in this non-relativistic limit has no direct connection to the energy. In analogy to the binding energy we may define a mass defect as

$$\Delta M_N = \frac{M_{\text{eff}}(N) - N M_{\text{eff}}(1)}{N} \quad . \tag{4.40}$$

If the mesons were real particles in the N-particle state then we could define a mass defect in terms of their masses

$$\overline{\Delta M}_N = \frac{M_{\text{eff}}(N) - (NM + <N_S> m_S + <N_V> m_V)}{N} \tag{4.41}$$

which, compared to $\Delta M_N$, is shifted by $\overline{\Delta M}_1$. These are given in table 1 for our four parameter sets. Shifts of the order of the nucleon mass point again to the breakdown of the non-relativistic treatment of nucleons. One can also define an excess meson mass by

$$\Delta M_{S,V} = \frac{m_S \Delta N_S + m_V \Delta N_V}{N} \tag{4.42}$$

where

$$\Delta N_i = <N_i(N)> - N <N_i(1)> \quad i = S, V \tag{4.43}$$

are the meson excess in the N-particle system. Whereas the effective mass was calculated in first order the meson numbers are calculated to zeroth order. The results are plotted in fig. 7 for all parameter sets. The mass defect increases more rapidly with $N$ than the excess meson mass. The vector excess per particle, $\Delta N_V/N$, is almost zero for the sets $G_1$ and $G_2$ rising rapidly for $G_4$. As expected there are, on the average, much more scalar mesons present.

---

[5] The results obtained with trial action 2 resemble very closely those of trial action 3 whereas for trial action 4 the values of $\Delta S$ are much smaller leaving as major cancelling contributions the diagonal and non-diagonal parts of $E_F$.

Finally we estimate various sizes of the N-particle state, all of which are calculated to zeroth order using the intrinsic wave function derived from the underlying model Hamiltonian. In fig. 8 we show the interparticle distance which settles to about 2 fm for $G_1$ and, as expected, decreases to about 0.5 fm for $G_4$. We generalize the definition of $r_{cm}$ given for trial action 1 in section 3 to apply to the other actions with the extra fictitious particle attached to each nucleon. We take

$$r_{\text{cm}}^2 = \frac{1}{N} \sum_{i=1}^{N} < \left( \frac{M_Z \mathbf{Z} + m_y \mathbf{y}_i + M \mathbf{x}_i}{M_Z + m_y + M} - \mathbf{x}_i \right)^2 > \qquad (4.44)$$

where the nucleon feels the presence of all other nucleons through the $Z-$particle. For trial action 3 this follows the interparticle distance expect for small $N$ where the system is unstable with respect to two-particle break-up.

How one measures the extension of the meson cloud around one bare nucleon is not very well defined since the $Z-$ particle is shared among all the nucleons. A lower bound is given by considering just the $y-$particle i.e. the expectation value of $(\mathbf{x}_i - \mathbf{y}_i)^2$. For trial action 4 this represents the only self energy contribution and it was the main motivation for introducing this trial action. As it can be seen this cloud radius is constant and small. An upper bound is given by summing both the contribution from the $y-$ and the $Z-$ particles i.e $< (\mathbf{Z} - \mathbf{x}_i)^2 + (\mathbf{x}_i - \mathbf{y}_i)^2 >$. The latter follows closely the interparticle distance.

## 5  Summary and Conclusions

We have presented a variational calculation of the N-particle bound state where dynamical meson exchange is summed up to all orders. We have for the first time included all vertex corrections and self energies in an $N$ particle system albeit in an approximate way. While the scalar and vector mesons are treated fully relativistically the $N$ constituents are treated non-relativistically. The main advantage of this method is that we are able to obtain analytic expressions for the observables of the N-body bound state and therefore study the behaviour as a function of $N$.

The crucial step in this work as in all variational approaches is the construction of a good trial action. For this we used the original idea by Feynman of replacing mesonic degrees of freedom by fictitious particles. This study revealed that it is of great importance to separate direct and exchange contributions giving them different strengths. This turns out to be much more important than distinguishing vector and scalar contributions. Studying trial action 4 we learn that although the strength of the purely exchange term in this action is small compared to that of the direct it significantly lowers the energy bound obtained with trial action 1 which makes no distinction between the self-energies and exchanges.

Using the Walecka model as our underlying theory requires the introduction of form factors. For the typical couplings of this model one has to tune the form factors to avoid large self energy contributions. We found that for a particular set of parameters we approximated a saturating system where the exchanged terms contribute a constant to the binding energy per particle cancelling to a large extent with the diagonal part. For these large cancelations occurring in a nearly saturating system it is important to construct a good trial action. This becomes less of an issue for a strongly bound system where the exchange contributions dominate. Another important feature that emerges from this calculation is that the number of excess mesons per constituent in a saturating system is rather small being practically zero as far as vector mesons are concerned and reaching about two for scalar mesons when $N \sim 100$. The radius of the dressed constituents defined solely in terms of the self energy comes out rather small and independent of the amount of binding. In contrast the interparticle distance is more than five times larger for the saturating system becoming almost equal to the cloud radius in the large binding situation.

Although the above features can not be taken as a quantitative description of the real nuclear case due to the lack of antisymmetrisation, nevertheless they are useful in that they reveal important difference between a saturating and a strongly bound system. In order to approach the nuclear system, in addition to the issue of antisymmetry, one has to treat the $\pi - N$ interaction which is strongly spin and isospin dependent. Treating the more realistic Chew-Low model [16] using coherent spin states [17] gives rise to a complex effective action [18] requiring a better understanding before variational results of similar accuracy as those obtained in the polaron problem can be produced. There has been some work [19] recently on the treatment of spin in the particle representation but it remains an open question whether we can apply those ideas within the present framework.

**Acknowledgement:** We thank R. Rosenfelder for many illuminating discussions and for a careful reading of the manuscript.

# A  Determination of the static effective potential

We derive here the static two-body potential from the effective action $S_\text{eff}$. Setting $F_i(k) = 1$ in eq. (2.8), we can perform the momentum integration in terms of the modified Bessel functions [20] leading to the following expression for the effective action

$$\begin{aligned}
S_\text{eff} &= \int_0^\beta dt \bigg\{ \sum_{i=1}^N \frac{M}{2}\dot{\mathbf{x}}_i^2 \\
&\quad - \frac{\alpha_S}{2\pi} \int_0^\beta dt' \sum_{i,j=1}^N \frac{m_S}{\sqrt{|\mathbf{x}_i(t) - \mathbf{x}_j(t')|^2 + |t - t'|^2}} K_1\left(m_S\sqrt{|\mathbf{x}_i(t) - \mathbf{x}_j(t')|^2 + |t - t'|^2}\right) \\
&\quad - \frac{\alpha_V}{2\pi} \int_0^\beta dt' \sum_{i,j=1}^N \bigg[ m_V^2 \int_0^1 da\, K_0\left(m_V\sqrt{a^2|\mathbf{x}_i(t) - \mathbf{x}_j(t')|^2 + |t - t'|^2}\right) \\
&\quad - \frac{m_V \dot{\mathbf{x}}_{i,\perp}(t) \cdot \dot{\mathbf{x}}_{j,\perp}(t')}{\sqrt{|\mathbf{x}_i(t) - \mathbf{x}_j(t')|^2 + |t - t'|^2}} K_1\left(m_V\sqrt{|\mathbf{x}_i(t) - \mathbf{x}_j(t')|^2 + |t - t'|^2}\right) \bigg] \\
&\quad + \frac{\alpha_V}{2\pi} \sum_{\substack{i,j=1 \\ i\neq j}}^N \frac{1}{|\mathbf{x}_i(t) - \mathbf{x}_j(t)|} \bigg\} \quad .
\end{aligned} \tag{A.1}$$

where constant terms coming from tadpole diagrams are omitted since they are irrelevant for the two-body potential. Due to the exponential suppression in $\tau = t - t'$ the main contribution in (A.1) comes from trajectories with $\mathbf{x}_i(t) \approx \mathbf{x}_i(t') \approx \mathbf{x}_i(T)$ where $T = \frac{t+t'}{2}$. In this approximation and neglecting the term proportional to the transverse velocities we can perform the $\tau$ integration [20]. We get the following result for the non-diagonal part of the action

$$\begin{aligned}
S_\text{eff}^\text{ND} &= \int_0^\beta dT \bigg[ -\frac{\alpha_S}{\pi} \sum_{\substack{i,j=1 \\ i\neq j}}^N \sqrt{\frac{m_S \pi}{2|\mathbf{x}_i(T) - \mathbf{x}_j(T)|}}\, K_{1/2}(m_S|\mathbf{x}_i(T) - \mathbf{x}_j(T)|) \\
&\quad - \frac{\alpha_V}{\pi} \sum_{\substack{i,j=1 \\ i\neq j}}^N \sqrt{\frac{m_V \pi}{2|\mathbf{x}_i(T) - \mathbf{x}_j(T)|}}\, m_V \int_0^1 da\, K_{-1/2}(m_V a|\mathbf{x}_i(T) - \mathbf{x}_j(T)|) \\
&\quad + \frac{\alpha_V}{\pi} \sum_{\substack{i,j=1 \\ i\neq j}}^N \frac{1}{|\mathbf{x}_i(T) - \mathbf{x}_j(T)|} \bigg] \quad .
\end{aligned} \tag{A.2}$$

Using the relation $K_{\pm 1/2}(z) = \sqrt{\frac{\pi}{2z}} e^{-z}$, the static potential between a pair of nucleons reduces to the well known result

$$V(r) = \alpha_V \frac{e^{-rm_V}}{r} - \alpha_S \frac{e^{-rm_S}}{r} \quad ; \quad r = |\mathbf{x}_i - \mathbf{x}_j| \tag{A.3}$$

a superposition of a Yukawa attraction due to scalar meson exchange and a Yukawa repulsion due to vector meson exchange.

# B Calculation of the relevant observables for a general quadratic action

We give here the details for the calculation of the expectation values needed to extract physical observables. They are applicable to any action quadratic in the particle trajectories.

## B.1 Energy and effective mass

The main quantity which has to be calculated is the mean value $< e^{i\mathbf{k}\cdot(\mathbf{x}_i(\tau)-\mathbf{x}_j(\sigma))} >_{S_F}$ with respect to the trial action $S_F$. For this purpose we apply the stationary phase approximation to the action

$$S' = S - \int_0^\beta dt \sum_{a=1}^N \mathbf{f}_a(t) \cdot \mathbf{x}_a(t) \tag{B.1}$$

where the source $\mathbf{f}_a(t)$ is given by

$$\mathbf{f}_a(t) = i\mathbf{k}(\delta_{ai}\delta(t-\tau) - \delta_{aj}\delta(t-\sigma)) \quad. \tag{B.2}$$

Using the trajectories $\tilde{\mathbf{x}}_i(t)$ which make $S'$ stationary one easily obtains

$$< e^{i\mathbf{k}\cdot(\mathbf{x}_i(\tau)-\mathbf{x}_j(\sigma))} >_{S_F} = \exp\left[\frac{1}{2}\int_0^\beta dt \sum_{a=1}^N \mathbf{f}_a(t) \cdot \tilde{\mathbf{x}}_a(t)\right] \quad. \tag{B.3}$$

In this way the problem reduces to finding $\tilde{\mathbf{x}}_i$. The required mean value follows by inserting the resulting expression in the above formula. To simplify the calculations we work in Fourier space using the following decomposition for the trajectories of the particles

$$\tilde{\mathbf{x}}_i(t) = \frac{1}{2}\tilde{\mathbf{a}}_i^0 + \sum_{n=1}^\infty \tilde{\mathbf{a}}_i^n \cos(2\pi n \frac{t}{\beta}) + \sum_{n=1}^\infty \tilde{\mathbf{b}}_i^n \sin(2\pi n \frac{t}{\beta}) \tag{B.4}$$

where, when two indices are used, subscripts denote particle labels and superscripts the Fourier component. In this representation we get, for a general quadratic action, the following set of equations which the Fourier coefficients $\tilde{\mathbf{a}}_i^0, \tilde{\mathbf{a}}_i^n$ and $\tilde{\mathbf{b}}_i^n$ must satisfy in order to make the action $S'$ stationary (we suppress the tildes in the following)

$$\begin{aligned}
\lambda_n \mathbf{a}_m^n - \Lambda_n \mathbf{A}_n &= \mathbf{X}_m^n \\
\lambda_n \mathbf{b}_m^n - \Lambda_n \mathbf{B}_n &= \mathbf{Y}_m^n \qquad m = 1,..,N \\
\lambda_0 \mathbf{a}_m^0 - \Lambda_0 \mathbf{A}_0 &= \mathbf{W}_m^0
\end{aligned} \tag{B.5}$$

where

$$\begin{aligned}
\mathbf{X}_m^n &= \frac{2i\mathbf{k}}{M\beta}\left[\delta_{mi}\cos(2\pi n\frac{\tau}{\beta}) - \delta_{mj}\cos(2\pi n\frac{\sigma}{\beta})\right] \\
\mathbf{Y}_m^n &= \frac{2i\mathbf{k}}{M\beta}\left[\delta_{mi}\sin(2\pi n\frac{\tau}{\beta}) - \delta_{mj}\sin(2\pi n\frac{\sigma}{\beta})\right] \\
\mathbf{W}_m^0 &= \frac{i\mathbf{k}}{M\beta}[\delta_{mi} - \delta_{mj}] \\
\mathbf{A}_n &= \sum_{m=1}^N \mathbf{a}_m^n, \quad \mathbf{B}_n = \sum_{m=1}^N \mathbf{b}_m^n, \quad \mathbf{A}_0 = \sum_{m=1}^N \mathbf{a}_m^0
\end{aligned} \tag{B.6}$$

Due to translation invariance we are free to choose $\mathbf{A}_0 = 0$. The coefficients $\lambda_n, \Lambda_n, \lambda_0, \Lambda_0$ are in general rational functions of $n$ and their exact form depends on the form of the trial action we use. For trial action 3 and 4 we obtain respectively

$$\lambda_n = c^2 + \left(\frac{2\pi n}{\beta}\right)^2 \left[\frac{v_D^2 + (\frac{2\pi n}{\beta})^2}{w_D^2 + (\frac{2\pi n}{\beta})^2}\right] \quad ; \quad \Lambda_n = \frac{c^2 w^2}{N\left[w^2 + (\frac{2\pi n}{\beta})^2\right]} \quad ; \quad \lambda_0 = c^2 \qquad (B.7)$$

and

$$\lambda_n = c^2 + \left(\frac{2\pi n}{\beta}\right)^2 \left[\frac{v_D^2 + (\frac{2\pi n}{\beta})^2}{w_D^2 + (\frac{2\pi n}{\beta})^2} - \Lambda_n\right] \quad ; \quad \Lambda_n = \frac{c^2 w^2}{N\left[w^2 + (\frac{2\pi n}{\beta})^2\right]} \quad ; \quad \lambda_0 = c^2 \qquad (B.8)$$

with

$$v_D^2 = c_D^2 + w_D^2 \qquad c_D^2 = \frac{4C_D}{Mw_D}$$
$$v^2 = c^2 + w^2 \qquad c^2 = \frac{4CN}{Mw} \qquad (B.9)$$

The expressions for trial actions 1 and 2 can be obtained from those of trial action 3 by setting $c_D = 0$ and $w = w_D$ respectively.

The system of linear equations given in eq. (B.5) has the solution

$$\begin{aligned} \mathbf{a}_m^n &= \frac{\mathbf{X}_m^n}{\lambda_n} + \frac{\Lambda_n \mathbf{X}_m^n}{\lambda_n(\lambda_n - N\Lambda_n)} \\ \mathbf{b}_m^n &= \frac{\mathbf{Y}_m^n}{\lambda_n} + \frac{\Lambda_n \mathbf{Y}_m^n}{\lambda_n(\lambda_n - N\Lambda_n)} \\ \mathbf{a}_m^0 &= \frac{\mathbf{W}_m^0}{\lambda_0} \end{aligned} \qquad (B.10)$$

Substituting in (B.4) these stationary values we obtain from (B.3)

$$< e^{i\mathbf{k}\cdot(\mathbf{x}_i(\tau) - \mathbf{x}_j(\sigma))} >_{S_F} = e^{\frac{-k^2}{2M}\mu_{ij}^2(\tau,\sigma)} \qquad (B.11)$$

where

$$\mu_{ij}^2(\tau,\sigma) = \frac{4}{N}[\Sigma_1(0)(N-1) - \Sigma_1(\tau-\sigma)(N\delta_{ij} - 1) + \Sigma_2(0) - \Sigma_2(\tau-\sigma)] \qquad (B.12)$$

valid in the limit $\beta \to \infty$. We have define the sums $\Sigma_1$ and $\Sigma_2$ by

$$\begin{aligned} \Sigma_1(t) &= \lim_{\beta \to \infty} \frac{1}{\beta} \sum_{n=1}^{\infty} \frac{\cos(\frac{2\pi nt}{\beta})}{\lambda_n} \\ \Sigma_2(t) &= \lim_{\beta \to \infty} \frac{1}{\beta} \sum_{n=1}^{\infty} \frac{\cos(\frac{2\pi nt}{\beta})}{\lambda_n - N\Lambda_n} \end{aligned}$$

In order to perform these sums we have to first find the roots in the numerator of these expressions and decompose the corresponding rational functions of $n$ in partial fractions. From this point on we will consider quadratic actions with any sum of exponential retardation factors. In this case the numerator $\lambda_n$ or $\lambda_n - N\Lambda_n$ can always be written as a real polynomial $P$ in $n^2$ of degree $p$ and the sums for the corresponding $\Sigma_j$ in the above expressions are given by

$$\Sigma_j(t) = \lim_{\beta \to \infty} \frac{1}{\beta} \sum_{n=1}^{\infty} \cos(\frac{2\pi n t}{\beta}) \sum_{l=1}^{p} \frac{C_{j,l}}{(\frac{2\pi n}{\beta})^2 + \rho_{j,l}^2} \tag{B.13}$$

where $-\rho_{j,l}^2$ are complex numbers denoting the roots of $P$ and $C_{j,l}$ are complex coefficients resulting from the partial fraction decomposition. To determine the contribution to the sum $\Sigma_j$ coming from a conjugate pair of roots we have to be able to calculate the following types of sums

$$T_1 = \sum_{n=1}^{\infty} \frac{\cos(\frac{2\pi n t}{\beta})}{\left((\frac{2\pi n}{\beta})^2 + x^2\right)^2 + y^4} \quad \text{and} \quad T_2 = \sum_{n=1}^{\infty} \frac{\left((\frac{2\pi n}{\beta})^2 + x^2\right)\cos(\frac{2\pi n t}{\beta})}{\left((\frac{2\pi n}{\beta})^2 + x^2\right)^2 + y^4} \quad . \tag{B.14}$$

Since we are interested in the limit $\beta \to \infty$ of the above expressions we can replace the sums by integrals and use the integration formulas of [20] to obtain $T_1$ and $T_2$. Having determined the sums $\Sigma_j(t)$ and consequently the mean value $< e^{i\mathbf{k}\cdot(\mathbf{x}_i(\tau)-\mathbf{x}_j(\sigma))} >_{S_F}$ we can repeat the above procedure using a suitable source term (see [3] for details) to evaluate the mean value $< \dot{\mathbf{x}}_{\perp,i}(\tau) \cdot \dot{\mathbf{x}}_{\perp,j}(\sigma) \, e^{i\mathbf{k}\cdot(\mathbf{x}_i(\tau)-\mathbf{x}_j(\sigma))} >_{S_F}$. The rest of the expectation values needed in (3.21) can be obtained from (B.3) by suitable differentiation with respect to $\mathbf{k}$. It is then straightforward to obtain also the effective mass using the definition of (3.22). The results for the energy and the effective mass are

$$E_N = E_F - < S_F >_{S_F} - \frac{1}{\pi} \int_0^{\beta} d\tau \int_0^{\infty} dk k^2 \left\{ N R_S(k,\tau) \left( (N-1) e^{-\frac{k^2}{2M}\mu^2(\tau)} + e^{-\frac{k^2}{2M}\mu_D^2(\tau)} \right) \right.$$
$$\left. + N R_V(k,\tau) \left( (N-1) \left[ \frac{m_V^2}{k^2} - \frac{G(\tau) - H(\tau)}{MN} \right] e^{-\frac{k^2}{2M}\mu^2(\tau)} + \left[ \frac{m_V^2}{k^2} + \frac{(N-1)G(\tau) + H(\tau)}{MN} \right] e^{-\frac{k^2}{2M}\mu_D^2(\tau)} \right) \right\}$$
$$+ \frac{1}{\pi} \int_0^{\infty} dk N(N-1) R_V(k,0) \omega_V e^{-\frac{k^2}{2M}\Sigma_1(0)} + C_0 \tag{B.15}$$

$$M_{\text{eff}} = NM \left( 1 + \frac{2}{3\pi} \int_0^{\beta} d\tau \int_0^{\infty} dk k^2 \left\{ \frac{k^2 \tau^2}{2M} R_S(k,\tau) \left( (N-1) e^{-\frac{k^2}{2M}\mu^2(\tau)} + e^{-\frac{k^2}{2M}\mu_D^2(\tau)} \right) \right. \right.$$
$$+ \frac{k^2 \tau^2}{2M} R_V(k,\tau) \left( (N-1) \left[ \frac{m_V^2}{k^2} - \frac{G(\tau) - H(\tau)}{NM} \right] e^{-\frac{k^2}{2M}\mu^2(\tau)} + \left[ \frac{m_V^2}{k^2} + \frac{(N-1)G(\tau) + H(\tau)}{NM} \right] e^{-\frac{k^2}{2M}\mu_D^2(\tau)} \right)$$
$$\left. \left. + \frac{2}{M} R_V(k,\tau) \left( (N-1) e^{-\frac{k^2}{2M}\mu^2(\tau)} + e^{-\frac{k^2}{2M}\mu_D^2(\tau)} \right) \right\} \right) \tag{B.16}$$

where we used the notation

$$\mu_{ij}^2(\tau,\sigma) = \begin{cases} \mu_D^2(\tau - \sigma) & \text{if } i = j \\ \mu^2(\tau - \sigma) & \text{if } i \neq j \end{cases} \tag{B.17}$$

The functions $G(\tau)$ and $H(\tau)$ are given by the second derivative of $\Sigma_1(\tau)$ and $\Sigma_2(\tau)$. The constant $C_0$ is given by eq.(3.28) and is independent of the variational parameters. The expressions for $E_0$ and $< S_F >_{S_F}$ in (B.15) depend on the exact form of the trial action and they are given explicitly below for trial action 3. The quantities $R_i(k,\tau)$ are defined in eq.(2.9).

In general the calculation of $\Sigma_1(t)$, $\Sigma_2(t)$ can only be accomplished numerically. For trial action 3 however the polynomial $P$ reduces to a biquadratic form and we can perform the calculations analytically to obtain for the energy of the $N$-particle system

$$\begin{aligned}
E_N = E_F &- \frac{1}{\pi} \int_0^\beta d\tau \int_0^\infty dk k^2 \left\{ N R_S(k,\tau) \left( (N-1) e^{-\frac{k^2}{2M}\mu^2(\tau)} + e^{-\frac{k^2}{2M}\mu_D^2(\tau)} \right) \right. \\
&+ N R_V(k,\tau) \left( (N-1) \left[ \frac{m_V^2}{k^2} - \frac{G(\tau) - H(\tau)}{MN} \right] e^{-\frac{k^2}{2M}\mu^2(\tau)} + \left[ \frac{m_V^2}{k^2} + \frac{(N-1)G(\tau) + H(\tau)}{MN} \right] e^{-\frac{k^2}{2M}\mu_D^2(\tau)} \right) \right\} \\
&+ \frac{1}{\pi} \int_0^\infty dk k^2 R_V(k,0) \left[ \frac{(\Omega_+ + \Omega_- - 1)(N-1)}{M} + N(N-1) \left( \frac{1 - \Omega_+ - \Omega_-}{NM} + \frac{\omega_V}{k^2} \right) e^{-\frac{k^2}{2M}(\frac{\Omega_+}{\eta_+} + \frac{\Omega_-}{\eta_-})} \right] \\
&+ C_0 - \frac{3}{4} c^2 \left[ (N-1) \left( \frac{\Omega_+}{\eta_+} + \frac{\Omega_-}{\eta_-} \right) + \frac{1}{w} \left( 1 - \frac{\Phi_+}{\zeta_+(w + \zeta_+)} - \frac{\Phi_-}{\zeta_-(w + \zeta_-)} \right) \right] \\
&- \frac{3}{4} c_D^2 \left[ (N-1) \left( \frac{\Omega_+}{w_D + \eta_+} + \frac{\Omega_-}{w_D + \eta_-} \right) + \frac{1}{w_D} \left( 1 - \frac{\Phi_+}{\zeta_+(w_D + \zeta_+)} - \frac{\Phi_-}{\zeta_-(w_D + \zeta_-)} \right) \right]
\end{aligned}$$
(B.18)

and the mass

$$\begin{aligned}
M_N = NM &\left( 1 + \frac{2}{3\pi} \int_0^\beta d\tau \int_0^\infty dk k^2 \left\{ \frac{k^2 \tau^2}{2M} R_S(k,\tau) \left( (N-1) e^{-\frac{k^2}{2M}\mu^2(\tau)} + e^{-\frac{k^2}{2M}\mu_D^2(\tau)} \right) \right. \right. \\
&+ \frac{k^2 \tau^2}{2M} R_V(k,\tau) \left( (N-1) \left[ \frac{m_V^2}{k^2} - \frac{G(\tau) - H(\tau)}{NM} \right] e^{-\frac{k^2}{2M}\mu^2(\tau)} + \left[ \frac{m_V^2}{k^2} + \frac{(N-1)G(\tau) + H(\tau)}{NM} \right] e^{-\frac{k^2}{2M}\mu_D^2(\tau)} \right) \\
&+ \frac{2}{M} R_V(k,\tau) \left( (N-1) e^{-\frac{k^2}{2M}\mu^2(\tau)} + e^{-\frac{k^2}{2M}\mu_D^2(\tau)} \right) \right\} \Bigg) \quad .
\end{aligned}$$
(B.19)

The quantities $\mu^2(\tau)$ and $\mu_D^2(\tau)$ are given by

$$\begin{aligned}
\mu^2(\tau) &= \frac{\Omega_-}{\eta_-} + \frac{\Omega_+}{\eta_+} - \frac{1}{N} \left[ \frac{\Omega_+}{\eta_+} \left( 1 - e^{-\eta_+ \tau} \right) + \frac{\Omega_-}{\eta_-} \left( 1 - e^{-\eta_- \tau} \right) \right] \\
&+ \frac{1}{N} \left[ \tau \left( 1 - \frac{\Phi_+}{\zeta_+^2} - \frac{\Phi_-}{\zeta_-^2} \right) + \frac{\Phi_+}{\zeta_+^3} \left( 1 - e^{-\zeta_+ \tau} \right) + \frac{\Phi_-}{\zeta_-^3} \left( 1 - e^{-\zeta_- \tau} \right) \right] \\
\mu_D^2(\tau) &= \frac{N-1}{N} \left[ \frac{\Omega_+}{\eta_+} \left( 1 - e^{-\eta_+ \tau} \right) + \frac{\Omega_-}{\eta_-} \left( 1 - e^{-\eta_- \tau} \right) \right] \\
&+ \frac{1}{N} \left[ \tau \left( 1 - \frac{\Phi_+}{\zeta_+^2} - \frac{\Phi_-}{\zeta_-^2} \right) + \frac{\Phi_+}{\zeta_+^3} \left( 1 - e^{-\zeta_+ \tau} \right) + \frac{\Phi_-}{\zeta_-^3} \left( 1 - e^{-\zeta_- \tau} \right) \right]
\end{aligned}$$
(B.20)

and the functions $G(\tau)$ and $H(\tau)$ by

$$\begin{aligned}
G(\tau) &= \Omega_+ \eta_+ \, e^{-\eta_+ \tau} + \Omega_- \eta_- \, e^{-\eta_- \tau} \\
H(\tau) &= \frac{\Phi_+}{\zeta_+} e^{-\zeta_+ \tau} + \frac{\Phi_-}{\zeta_-} e^{-\zeta_- \tau} \quad .
\end{aligned}$$
(B.21)

The quantities $\Omega, \eta, \Phi, \zeta$ are determined in terms of the variational parameters through

$$\Omega_\pm = \frac{1}{2} \left( 1 \pm \frac{c^2 + c_D^2 - w_D^2}{2\sqrt{\left(\frac{v_D^2 - c^2}{2}\right)^2 + c^2 c_D^2}} \right)$$

$$\eta_\pm = \sqrt{\frac{v_D^2+c^2}{2} \pm \sqrt{\left(\frac{v_D^2-c^2}{2}\right)^2 + c^2 c_D^2}}$$

$$\Phi_\pm = \frac{1}{2}\left(c_D^2 + c^2 \pm \frac{(c^2+c_D^2)^2 + (c^2-c_D^2)(w^2-w_D^2)}{2\sqrt{\left(\frac{v_D^2-v^2}{2}\right)^2 + c^2 c_D^2}}\right)$$

$$\zeta_\pm = \sqrt{\frac{v_D^2+v^2}{2} \pm \sqrt{\left(\frac{v_D^2-v^2}{2}\right)^2 + c^2 c_D^2}} \tag{B.22}$$

The ground state energy $E_F$ for trial action 3 reads

$$E_F = \frac{3}{2}\left((N-1)\sqrt{v_D^2 + c^2 + 2cw_D} + \sqrt{v_D^2 + v^2 + 2\sqrt{v^2 v_D^2 - c^2 c_D^2}} - Nw_D - w\right) \tag{B.23}$$

which vanishes in the free case.

We can again recover the results for trial action 1 by taking $c_D \to 0$ and for trial action 2 by taking $c \to 0$ in the above formulae. For trial action 4 these expressions for the energy and effective are evaluated numerically.

## B.2 Radii

The mean inter-particle distance $\bar{r}_N$ is given in general by

$$<\bar{r}_N^2> = \frac{1}{N(N-1)} \sum_{i,j=1}^N <(\mathbf{x}_i - \mathbf{x}_j)>^2 = \frac{3}{2M}\Sigma_1(0) \tag{B.24}$$

The calculation of the cloud and rms radius is in general more complicated and the resulting expressions depend on the form of the trial action. In general one has to resort to a numerical evaluation. We describe here the calculation of these quantities in the case of trial action 3. To define the rms or center of mass radius, $r_{\rm cm}$, we consider the subsystem of the $i$th nucleon, the $Z-$ and the $y_i$ particles. We take the distance of the nucleon from the center of mass of this subsystem to be the rms radius

$$<r_{\rm cm}^2> = \frac{1}{N} \sum_{i=1}^N <\left(\frac{M_Z \mathbf{Z} + m_y \mathbf{y}_i + M\mathbf{x}_i}{M_Z + m_y + M} - \mathbf{x}_i\right)^2> . \tag{B.25}$$

The cloud radius should give a measure of the extension of the meson cloud around the bare nucleon and therefore one should sum up the all the self energy contributions. For the single particle system this is given by $<(\mathbf{Z}-\mathbf{x})^2 + (\mathbf{y}-\mathbf{x})^2>^{1/2}$ when a $y$-particle is present. For more than one particle the $Z$-particle does no longer correspond to a purely self energy contribution but models also the exchange of mesons between different nucleons. Since the $y$-particles describe purely the self energies we may take

$$r_{\rm cloud}^2 = <r_y^2> = \frac{1}{N} \sum_{i=1}^N <(\mathbf{y}_i - \mathbf{x}_i)^2> \tag{B.26}$$

as a lower bound to the cloud radius. The quantity

$$\tilde{r}^2_{\text{cloud}} = <r_y^2 + r_Z^2> \quad ; \quad <r_Z^2> = \frac{1}{N}\sum_{i=1}^{N} < (\mathbf{Z} - \mathbf{x}_i)^2 > \tag{B.27}$$

greatly overestimates the dressing since it probes via meson exchange also the interparticle separation. In trial action 4 the contribution of the $Z-$ particle to the self energy is turned off by hand by simply not including the corresponding diagonal term. Although for this action the cloud radius is given entirely by $<r_y^2>^{1/2}$ we do not have the wave function since this action is not described by a model Lagrangian. Using the wave function of trial action 3 with the variational parameters derived for 4 is the approximation which we adopt.

To calculate the expectation values in the above equations we find the wave function of the system described through the Langrangian (3.17). The easiest way to do this is to go over to normal mode coordinates. For this we use the scaled variables $\mathbf{x}'_i = \sqrt{M}\mathbf{x}_i$, $\mathbf{y}'_i = \sqrt{m_y}\mathbf{y}_i$ and $\mathbf{Z}' = \sqrt{M_Z}\mathbf{Z}$ and we express the potential energy part of (3.15) in terms of these variables. The corresponding potential energy matrix for trial action 3 is given by

$$\hat{V} = \begin{pmatrix}
\frac{K+\kappa}{M} & 0 & \cdots & 0 & \frac{-\kappa}{\sqrt{Mm_y}} & 0 & \cdots & 0 & \frac{-K}{\sqrt{MM_Z}} \\
0 & \frac{K+\kappa}{M} & \cdots & 0 & 0 & \frac{-\kappa}{\sqrt{Mm_y}} & \cdots & 0 & \frac{-K}{\sqrt{MM_Z}} \\
\vdots & \vdots & \vdots & \vdots & \vdots & \vdots & \vdots & \vdots & \vdots \\
0 & 0 & \cdots & \frac{K+\kappa}{M} & 0 & 0 & \cdots & \frac{-\kappa}{\sqrt{Mm_y}} & \frac{-K}{\sqrt{MM_Z}} \\
\frac{-\kappa}{\sqrt{Mm_y}} & 0 & \cdots & 0 & \frac{\kappa}{m_y} & 0 & \cdots & 0 & 0 \\
0 & \frac{-\kappa}{\sqrt{Mm_y}} & \cdots & 0 & 0 & \frac{\kappa}{m_y} & \cdots & 0 & 0 \\
\vdots & \vdots & \vdots & \vdots & \vdots & \vdots & \vdots & \vdots & \vdots \\
0 & 0 & \cdots & \frac{-\kappa}{\sqrt{Mm_y}} & 0 & 0 & \cdots & \frac{\kappa}{m_y} & 0 \\
\frac{-K}{\sqrt{MM_Z}} & \frac{-K}{\sqrt{MM_Z}} & \cdots & \frac{-K}{\sqrt{MM_Z}} & 0 & 0 & \cdots & 0 & \frac{NK}{M_Z}
\end{pmatrix} \tag{B.28}$$

We then determine numerically the matrix $A$ which brings $\hat{V}$ into a diagonal form. In terms of this matrix the coordinates $\mathbf{x}'_i, \mathbf{y}'_i, \mathbf{Z}'$ are given by

$$\mathbf{x}'_i = \sum_{j=1}^{2N+1} A(i,j)\mathbf{q}_j, \quad i = 1,..,N$$

$$\mathbf{y}'_i = \sum_{j=1}^{2N+1} A(i,j)\mathbf{q}_j, \quad i = N,..,2N$$

$$\mathbf{Z}'_i = \sum_{j=1}^{2N+1} A(2N+1,j)\mathbf{q}_j \tag{B.29}$$

where $\mathbf{q}_i$, $i = 1,..,2N+1$ are the normal mode coordinates of the system. In these coordinates the wave function of the system is given by

$$\Psi(\mathbf{q}_1,...,\mathbf{q}_{2N+1}) = \mathcal{N}\exp\left[-\frac{1}{2}\sum_{i=1}^{2N+1}\epsilon_i \mathbf{q}_i^2\right] \tag{B.30}$$

where $\epsilon_i$ are the eigenvalues of the matrix $\hat{V}$. Due to translation invariance one of these eigenvalues corresponding to the center of mass vanishes. Without loss of generality we take the vanishing eigenvalue to be $\epsilon_1$. Then the expectation values in eq. (B.25-B.27) with respect to the wave function of (B.30) can be calculated without difficulty leading to the following expressions

$$\begin{aligned} <r_Z^2> &= \sum_{i=1}^{N}\sum_{j=2}^{2N+1} \frac{3}{2N\sqrt{\epsilon_j}} \left[\frac{A(2N+1,j)}{\sqrt{M_Z}} - \frac{A(i,j)}{\sqrt{M}}\right]^2 \\ <r_y^2> &= \sum_{i=1}^{N}\sum_{j=2}^{2N+1} \frac{3}{2N\sqrt{\epsilon_j}} \left[\frac{A(i+N,j)}{\sqrt{m_y}} - \frac{A(i,j)}{\sqrt{M}}\right]^2 \\ <r_{cloud}^2> &= <r_Z^2> + <r_y^2> \\ <r_{cm}^2> &= \sum_{i=1}^{N}\sum_{j=2}^{2N+1} \frac{3}{2N\sqrt{\epsilon_j}} \left[\frac{M_Z}{M_Z+m_y+M}\left(\frac{A(2N+1,j)}{\sqrt{M_Z}} - \frac{A(i,j)}{\sqrt{M}}\right)\right. \\ &\quad \left. + \frac{m_y}{M_Z+m_y+M}\left(\frac{A(i+N,j)}{\sqrt{m_y}} - \frac{A(i,j)}{\sqrt{M}}\right)\right]^2 \end{aligned} \quad (B.31)$$

In the case of trial action 1 the above expressions can be evaluated analytically leading to the formulae given in section 3.

**Figure captions**

**Figure 1** : The static potential of eq.(2.14) for five different values of the scalar and vector couplings and cutoffs. The dotted line corresponds to $G_0$ (infinite cut offs and $\alpha_S = 6.52$ and $\alpha_V = 10.84$), the long dash-dotted to $G_1$ ($\alpha_S = 6.52, \Lambda_S = 0.54$ GeV and $\alpha_V = 10.84, \Lambda_V = 0.92$ GeV), the long dashed to $G_2$ ($\alpha_S = 3.2, \Lambda_S = 1.0$ GeV and $\alpha_V = 5.12, \Lambda_V = 1.4$ GeV), the short dash-dotted to $G_3$ ($\alpha_S = 2.4, \Lambda_S = 1.0$ GeV and $\alpha_V = 4.0, \Lambda_V = 1.0$ GeV), and the short dashed to $G_4$ ($\alpha_S = 3.4, \Lambda_S = 1.0$ GeV and $\alpha_V = 3.57\ \Lambda_V = 1.2$ GeV).

**Figure 2** : The system of harmonic oscillators which describes the trial action of eq.(3.18).

**Figure 3** : The variational parameters of the trial actions 1 to 4 are shown in (i) to (iv) respectively. We use lines to show the parameters for the set $G_2$ and symbols for the set $G_4$. The purely diagonal variational parameters are scaled in both cases by a factor of 0.1. They are shown by the dotted line ( retardation $w_d/10$) and by the short dashed line ($v_d/10$), whereas the long dashed (retardation $w$) and dash-dotted ($v$) lines show variational parameters for the exchange terms. The corresponding diagonal parameters are shown by the triangles and crosses and the exchange by the squares and x's. In (iv) we do not plot the variational parameters for the set $G_4$ since their behaviour is the same as for the set $G_4$ in (iii).

**Figure 4** : The binding energy per particle in GeV versus the number of particles. For the parameter set $G_1$ trial action 1 produces no binding, trial action 3 and 4 are shown with the dash-dotted and solid lines respectively. For set $G_2$ trial action 1 produces no binding, trial action 3 and 4 are shown with the dashed and dotted lines respectively. The crosses and squares correspond to trial action 1 and 3 (same as trial action 4) for the set $G_3$ and the stars and triangles to trial action 1 and 3 (same as trial action 4) for the set $G_4$. The results for trial action 2 coincide in all cases with those from action 4.

**Figure 5** : The binding energy per nucleon as a function of $N$ for trial action 3 is shown by the dashed line. The squares and crosses show the diagonal and non-diagonal contributions.
(a) Parameter set $G_1$, (b) Parameter set $G_2$, (c) Parameter set $G_3$, (d) Parameter set $G_4$.

**Figure 6** : (a) The energy in GeV is shown versus the number of particles by the dashed line for the parameter set $G_1$ using trial action 3. The squares and triangles denote the diagonal contributions from $\Delta S$ and the trial energy respectively and the x's and crosses the corresponding non-diagonal parts. (b) The same as (a) but for the parameter set $G_4$.
(c) The same as (a) but for the binding energy per particle.
(d) The same as (c) but for the binding energy per particle.

**Figure 7** : The effective mass defect (triangles) and the excess meson mass (crosses) per particle in GeV as well as the scalar (dash-dotted line) and vector (dotted line) meson excess per particle are shown using trial action 3 for the different parameter sets. The (a)-(d) correspond to the same set of values as those denoted in figure 5.

**Figure 8** : The interparticle distance is shown with the dotted line and the center of mass radius with the squares both using trial action 3. The cloud radius is shown with the dashed line and it is obtained using the parameters of trial action 4. The notation (a)-(d) is the same as that in figure 5.

**Table 1:** We give the values of the scalar and vector couplings and cut offs, $g_S$, $g_V$, $\Lambda_S$ and $\Lambda_V$ respectively and denote a given set of these values by the symbol given in the first column. The rest of the columns give single particle properties which coincide when using trial actions 1,2 and 4. The last six columns give the values of the bare mass and the mass defect in GeV, the average number of scalar and vector mesons and the rms and cloud radius in fm.

|       | $\alpha_S$ | $\alpha_V$ | $\Lambda_S$ (GeV) | $\Lambda_V$ (GeV) | $M_0$ (GeV) | $\overline{\Delta M}_1$ (GeV) | $<N_S>$ | $<N_V>$ | $r_{\rm cm}$ (fm) | $r_{\rm cloud}$ (fm) |
|-------|------|-------|------|-------|-------|--------|------|------|------|------|
| $G_0$ | 6.52 | 10.84 | ∞    | ∞     |       |        |      |      |      |      |
| $G_1$ | 6.52 | 10.84 | 0.54 | 0.921 | 0.181 | -0.888 | 0.21 | 1.96 | 0.24 | 0.28 |
| $G_2$ | 3.2  | 5.12  | 1.0  | 1.4   | 0.272 | -1.193 | 0.21 | 2.23 | 0.18 | 0.24 |
| $G_3$ | 2.4  | 4.0   | 1.0  | 1.0   | 0.578 | -0.267 | 0.16 | 0.69 | 0.12 | 0.36 |
| $G_4$ | 3.4  | 3.57  | 1.0  | 1.2   | 0.517 | -0.469 | 0.22 | 0.99 | 0.13 | 0.30 |

**Table 2:** The values of the total energy $E_N^{(j)}$ and the binding energy $BE_N^{(j)}$ for the parameter set $G_2$ using the trial action $j, (j=1,..,4)$ are given for various particle numbers $N$.

| $N$ | $E_N^{(1)}$ (GeV) | $E_N^{(2)}$ (GeV) | $E_N^{(3)}$ (GeV) | $E_N^{(4)}$ (GeV) | $BE_N^{(1)}$ (GeV) | $BE_N^{(2)}$ (GeV) | $BE_N^{(3)}$ (GeV) | $BE_N^{(4)}$ (GeV) |
|-----|---------|----------|----------|----------|--------|-------|-------|-------|
| 1   | -1.790  | -1.790   | -1.838   | -1.790   | 0.0    | 0.0   | 0.0   | 0.0   |
| 2   | -2.482  | -3.656   | -3.741   | -3.677   | -0.549 | 0.038 | 0.032 | 0.049 |
| 4   | -4.423  | -7.311   | -7.482   | -7.325   | -0.684 | 0.038 | 0.032 | 0.042 |
| 6   | -6.431  | -10.975  | -11.222  | -10.987  | -0.718 | 0.040 | 0.032 | 0.042 |
| 8   | -8.485  | -14.649  | -14.963  | -14.646  | -0.730 | 0.041 | 0.032 | 0.041 |
| 10  | -10.549 | -18.338  | -18.727  | -18.335  | -0.735 | 0.044 | 0.035 | 0.044 |
| 20  | -21.273 | -37.088  | -37.872  | -37.130  | -0.726 | 0.065 | 0.055 | 0.067 |
| 30  | -32.545 | -56.359  | -57.541  | -56.402  | -0.705 | 0.089 | 0.080 | 0.091 |
| 40  | -44.352 | -76.157  | -77.739  | -76.192  | -0.681 | 0.114 | 0.105 | 0.115 |
| 50  | -56.691 | -96.483  | -98.467  | -96.535  | -0.656 | 0.140 | 0.131 | 0.141 |
| 100 | -126.333| -206.048 | -210.045 | -206.188 | -0.526 | 0.271 | 0.262 | 0.271 |

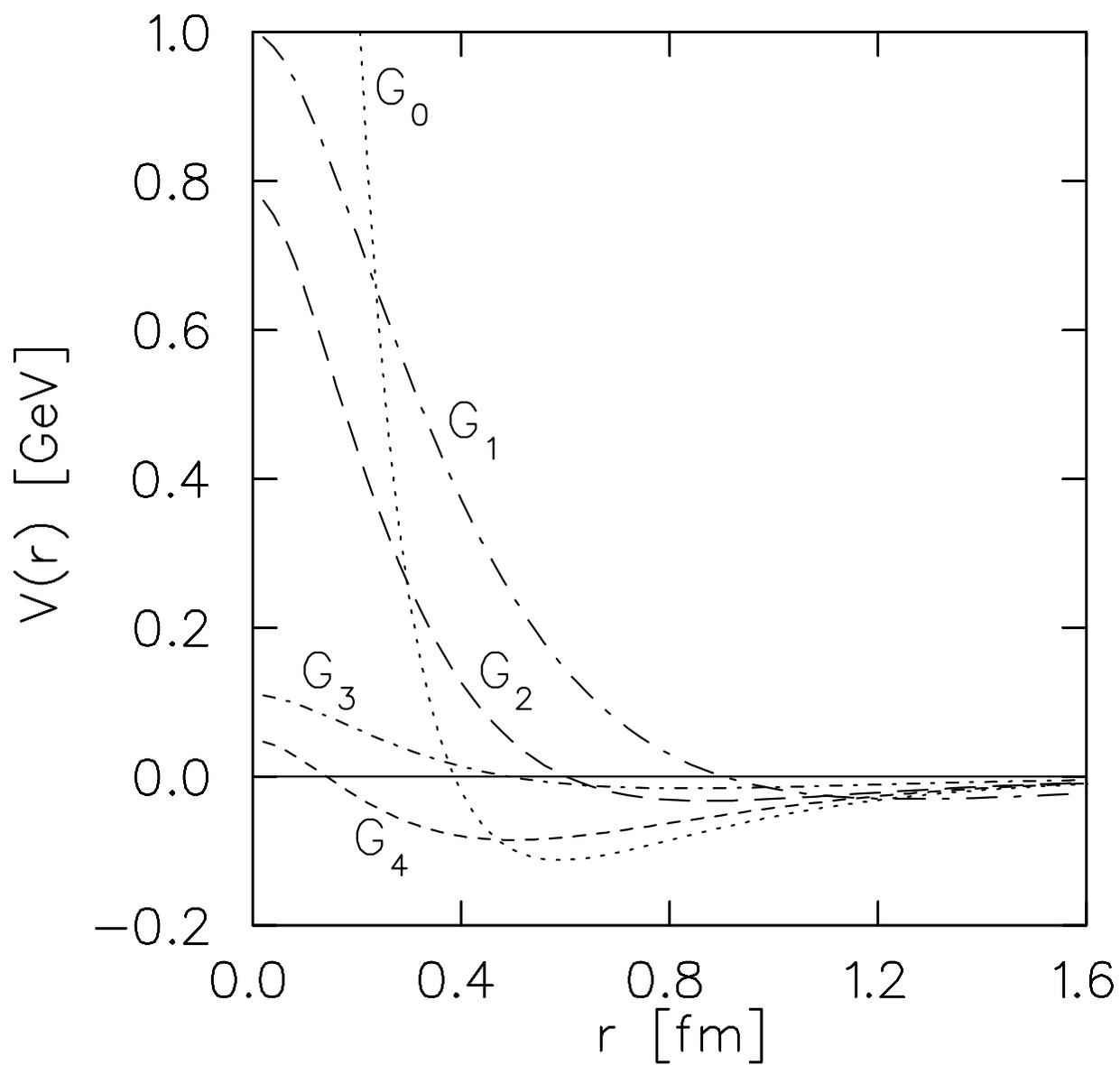

Figure 1

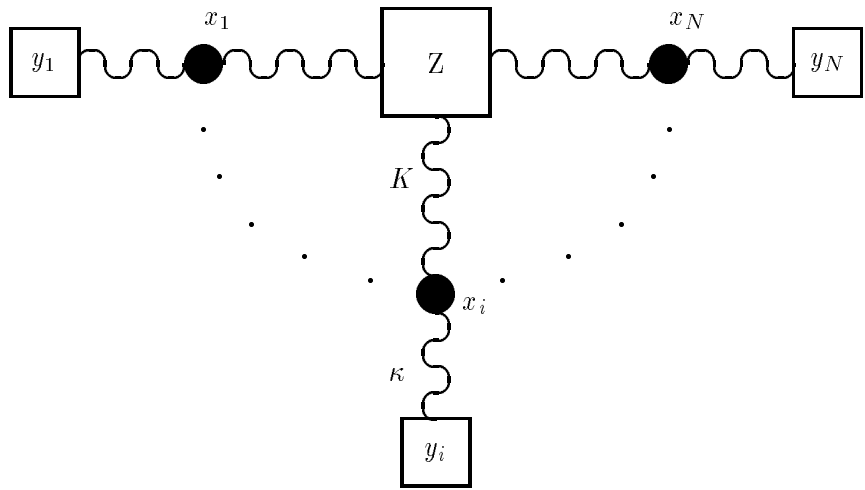

**Figure 2**

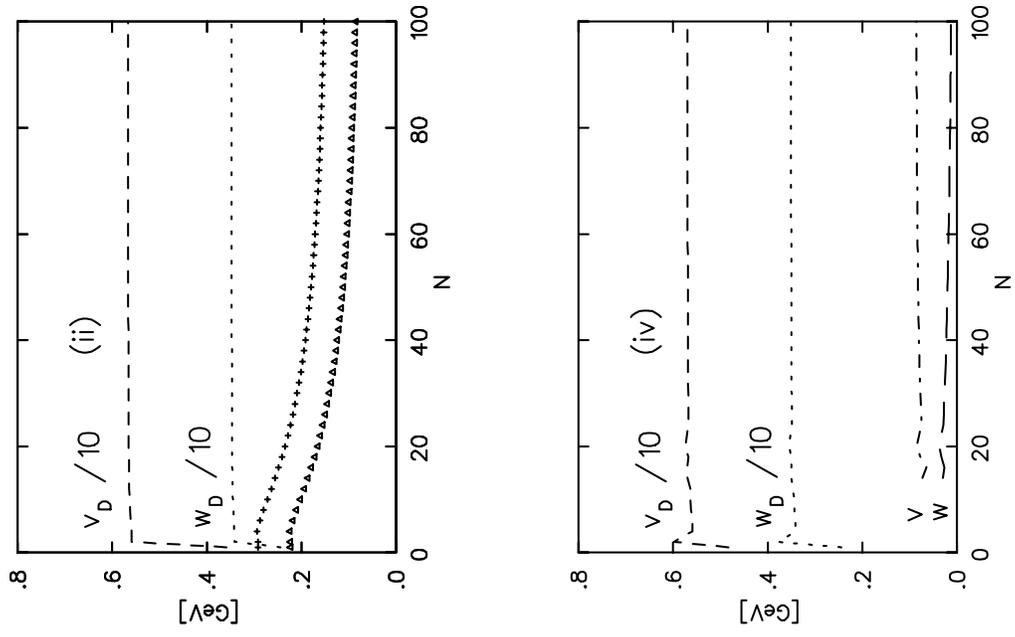

Figure 3

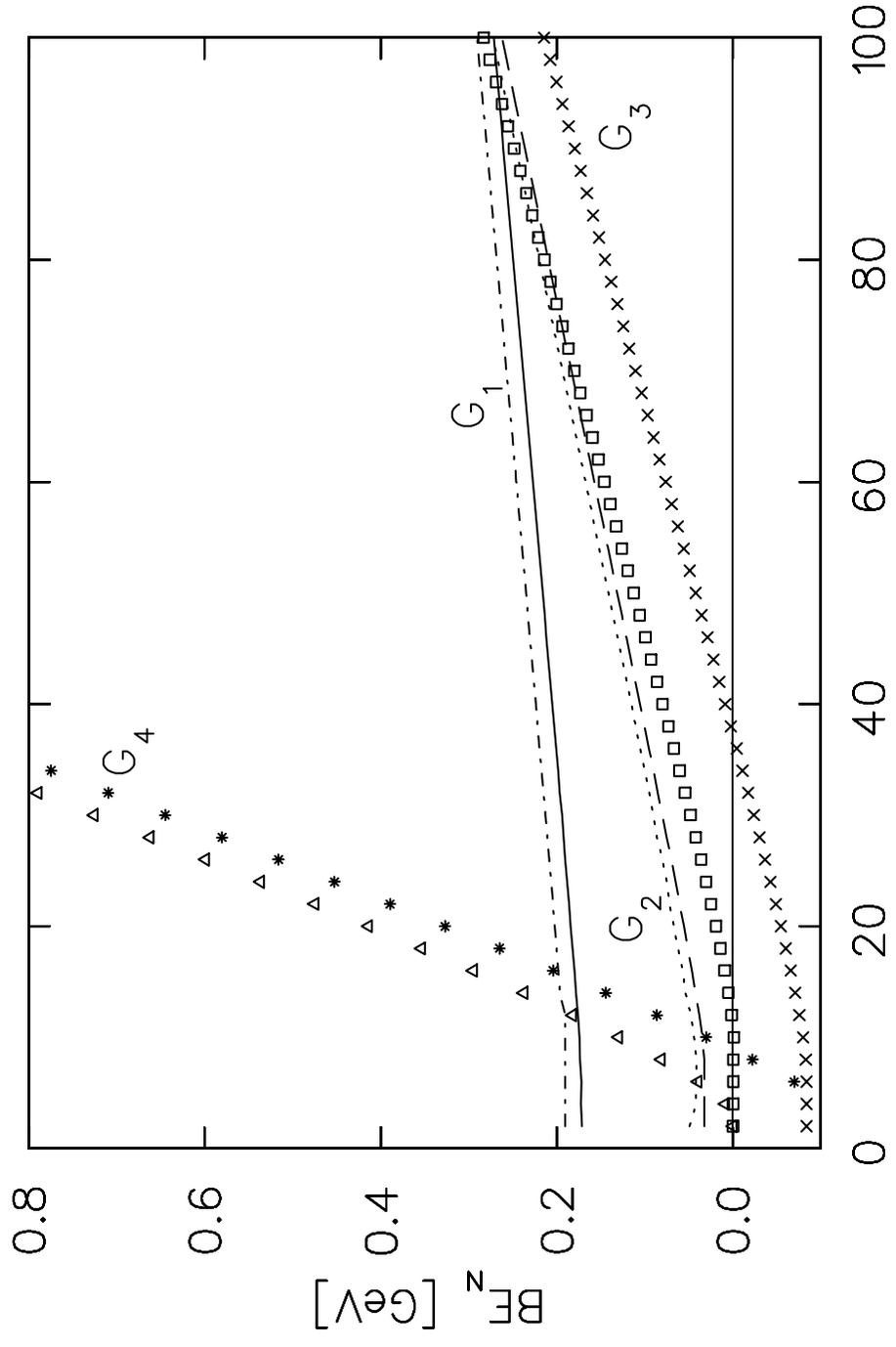

Figure 4

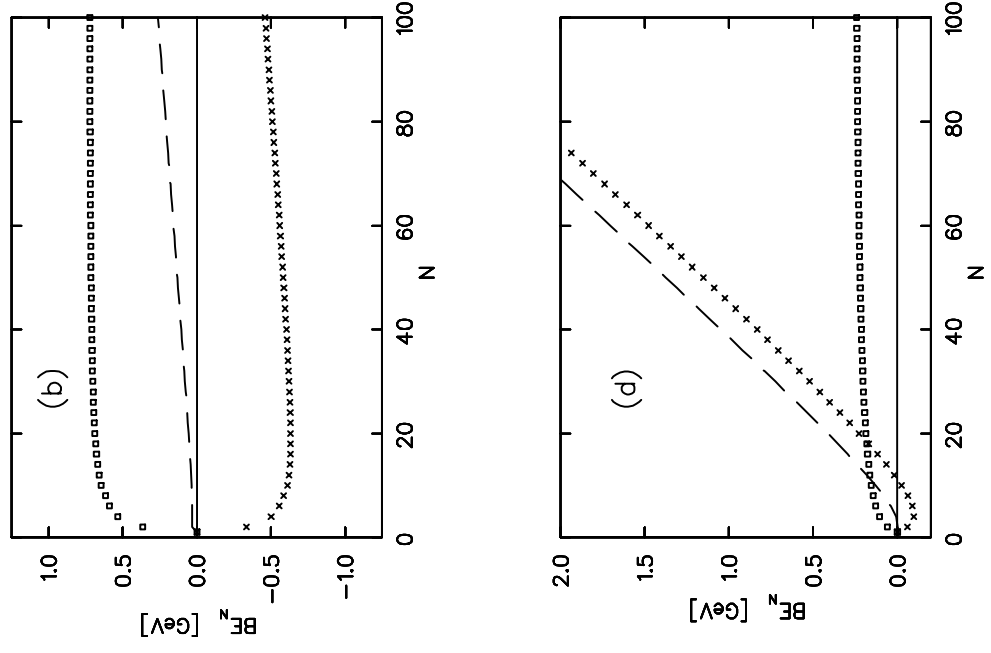

Figure 5

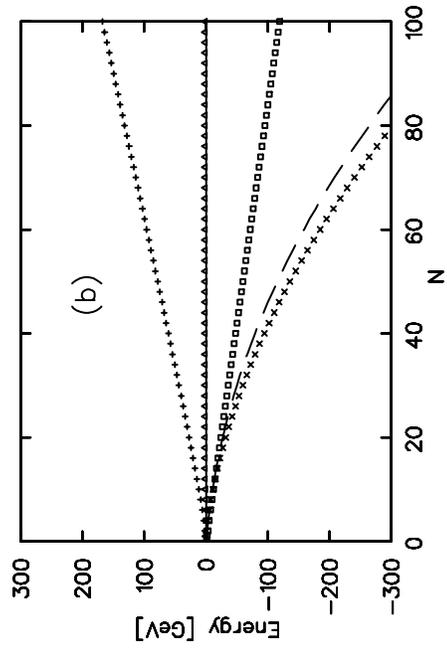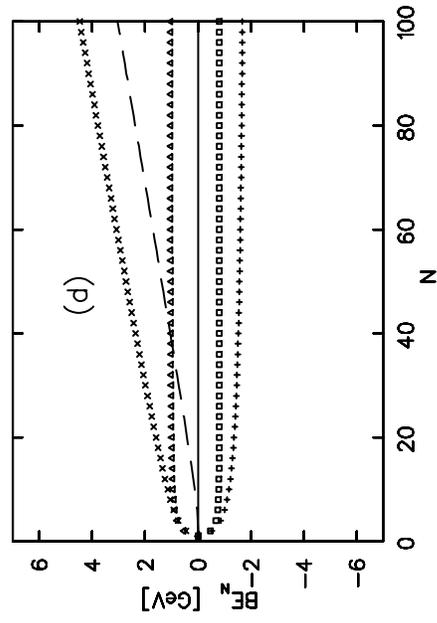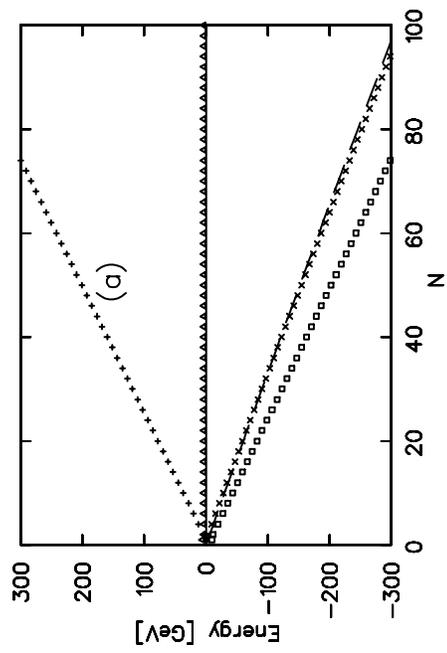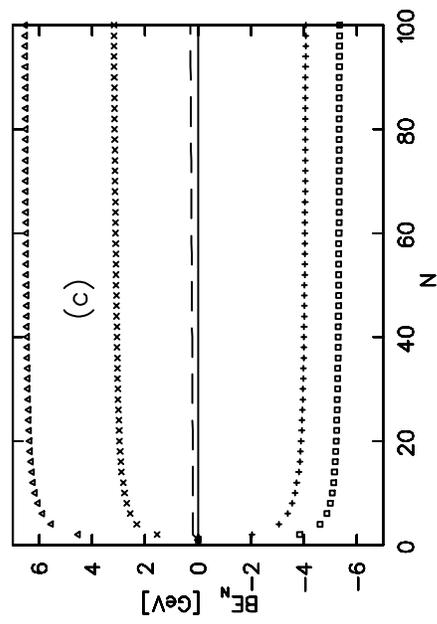

Figure 6

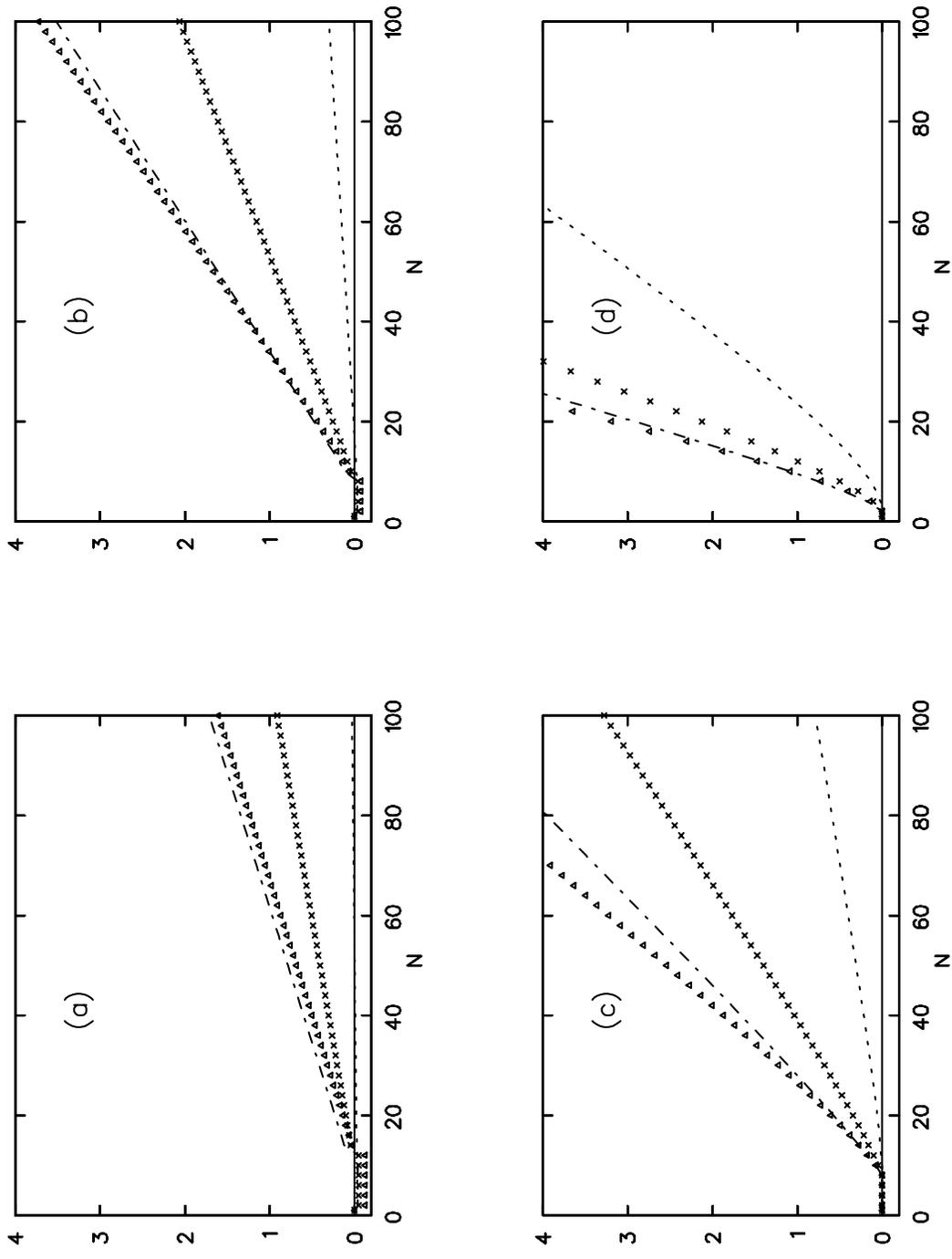

Figure 7

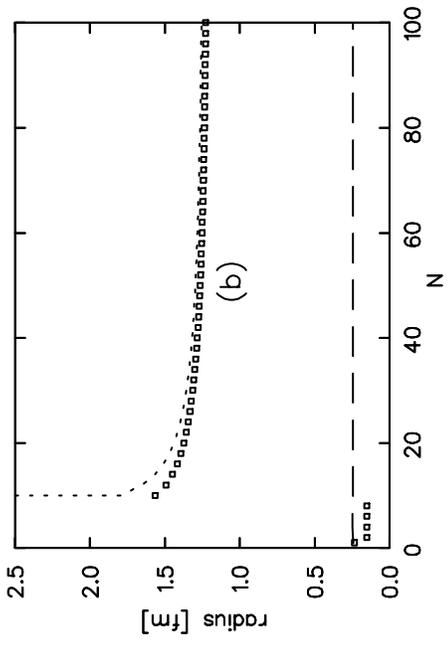
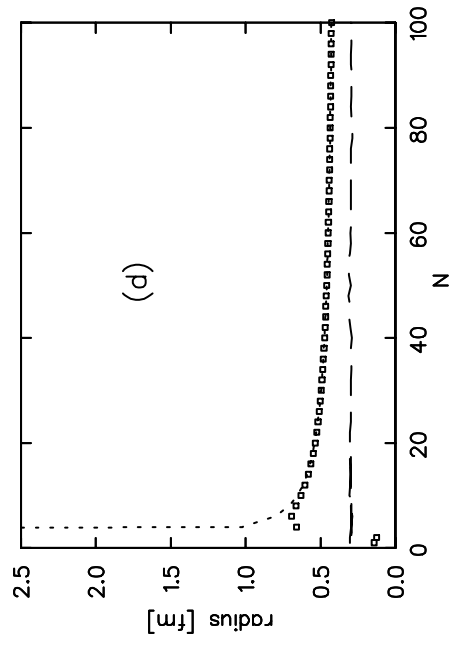
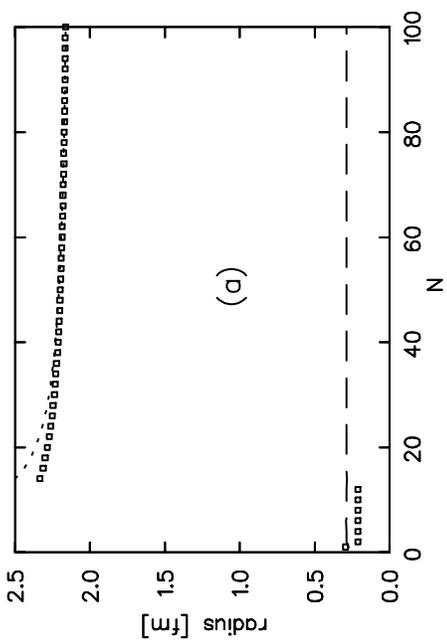
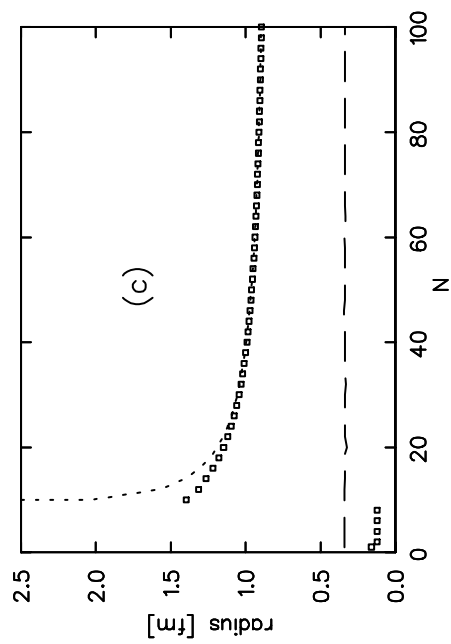

Figure 8